\documentclass[lettersize,journal]{IEEEtran}
\usepackage{amsmath,amsfonts}
\usepackage{algorithmic}
\usepackage{array}
\usepackage[caption=false,font=normalsize,labelfont=sf,textfont=sf]{subfig}
\usepackage{textcomp}
\usepackage{stfloats}
\usepackage{url}
\usepackage{verbatim}
\usepackage{graphicx}
\usepackage{booktabs}
\usepackage{graphicx}
\usepackage{multirow}
\usepackage{svg}
\usepackage{amssymb}
\usepackage{placeins}

\usepackage{bm}
\usepackage{xcolor}

\usepackage{tabularray}

\def\BibTeX{{\rm B\kern-.05em{\sc i\kern-.025em b}\kern-.08em
    T\kern-.1667em\lower.7ex\hbox{E}\kern-.125emX}}
\usepackage{balance}
\begin{document}
\title{Interleaved Block-based Learned Image Compression with Feature Enhancement and Quantization Error Compensation}
\author{ Shiqi Jiang, Hui Yuan, Senior Member, IEEE, Shuai Li, Senior Member, IEEE, Raouf Hamzaoui, Senior Member, IEEE, Xu Wang, Member, IEEE, and Junyan Huo, Member, IEEE
\thanks{This work was supported in part by the National Natural Science Foundation of China under Grants 62222110 and 62172259, the Taishan Scholar Project of Shandong Province (tsqn202103001), and the Natural Science Foundation of Shandong Province under Grant ZR2022ZD38, and the High-end Foreign Experts Recruitment Plan of Chinese Ministry of Science and Technology under Grant G2023150003L. (\textit{Corresponding author: Hui Yuan.})

Shiqi Jiang is with the School of software, Shandong University, Jinan, Shandong, China(e-mail: shiqijiang@mail.sdu.edu.cn).

Hui Yuan and Shuai Li are with the School of Control Science and Engineering, Shandong University, Jinan, Shandong, China (e-mail: huiyuan@sdu.edu.cn).

Raouf Hamzaoui is with the School of Engineering and Sustainable Development, De Montfort University, LE1 9BH Leicester, U.K.

Xu Wang is with the College of Computer Science and Software Engineering, Shenzhen University, Shenzhen, China.

Junyan Huo is with the School of Telecommunications Engineering, Xidian University, Xi'an, China.}}

\markboth{Journal of \LaTeX\ Class Files,~Vol.~18, No.~9, September~2020}%
{How to Use the IEEEtran \LaTeX \ Templates}

\maketitle

\begin{abstract}
In recent years, learned image compression (LIC) methods have achieved significant performance improvements. However, obtaining a more compact latent representation and reducing the impact of quantization errors remain key challenges in the field of LIC. To address these challenges, we propose a feature extraction module, a feature refinement module, and a feature enhancement module. Our feature extraction module shuffles the pixels in the image, splits the resulting image into sub-images, and extracts coarse features from the sub-images. Our feature refinement module stacks the coarse features and uses an attention refinement block composed of concatenated three-dimensional convolution residual blocks to learn more compact latent features by exploiting correlations across channels, within sub-images (intra-sub-image correlations), and across sub-images (inter-sub-image correlations). Our feature enhancement module reduces information loss in the decoded features following quantization. We also propose a quantization error compensation module that mitigates the quantization mismatch between training and testing. Our four modules can be readily integrated into state-of-the-art LIC methods. Experiments show that combining our modules with Tiny-LIC outperforms existing LIC methods and image compression standards in terms of peak signal-to-noise ratio (PSNR) and multi-scale structural similarity (MS-SSIM) on the Kodak dataset and the CLIC dataset.

\end{abstract}


\section{Introduction}
As one of the most important multimedia data types, images play an indispensable role in people's daily lives. To address the challenges of transmission and storage caused by the explosion of massive image data, image compression is required. In recent years, learned image compression (LIC) techniques have made significant progress \cite{1,2,3,9447950}, outperforming traditional methods in terms of rate-distortion (RD) performance. The latest LIC methods \cite{4,5} have surpassed the H.266/VVC \cite{8} standard, which is currently one of the best non-learning-based methods to compress images and videos in terms of peak signal-to-noise ratio (PSNR) and multi-scale structural similarity (MS-SSIM) metrics. This demonstrates the great potential of LIC methods for future applications.

LIC methods based on the variational autoencoder (VAE) \cite{balle2018variational} architecture have been widely applied, focusing on two aspects: how to learn more accurate entropy models and how to learn more compact latent representations. Early works \cite{9,balle2016end,agustsson2017soft,theis2017lossy} used relatively simple entropy models, which have limited performance. Currently, the mainstream approaches \cite{balle2018variational,cheng2020learned,guo2021causal,minnen2018joint,minnen2020channel,fu2023learned,he2022elic,10120973} use context-adaptive techniques to improve entropy models. Minninen \textit{et al}. \cite{minnen2018joint} combined an autoregressive model and a hyperprior model to obtain more accurate Gaussian distribution parameters for estimating latent feature distributions. However, this sequential approach greatly limits the encoding efficiency. He \textit{et al}. \cite{He_2021_CVPR} proposed a checkerboard context model to improve the entropy model in a parallel way, achieving a good balance between coding performance and time complexity. Inspired by parallel probability estimation research, Lu \textit{et al}. \cite{lu2022high} proposed a multi-stage context model (MCM), achieving an excellent performance-complexity trade-off.

As learned image compression can be formulated as a nonlinear transform coding problem, learning transformation features is a key factor in improving performance. Previous works introduced deeper networks \cite{gao2021neural,guo2021causal,lin2020spatial}, attention modules \cite{liu2020unified,liu2019non,cheng2020learned,fu2023learned}, or invertible structures \cite{ma2020end,xie2021enhanced} into VAE-based codecs. Although these structures can improve performance to some extent, the ability of feature extraction is still limited, i.e., redundant information still exists in the latent representation. To further improve the coding efficiency, we propose feature extraction, refinement, and enhancement modules. To mitigate the mismatch between training with soft quantization and testing with hard quantization, we also propose a quantization error compensation module. All our modules are trainable and can be easily integrated into state-of-the-art LIC methods as plug and play components. Our contributions are as follows. 

\begin{figure*}
	\centering
	\begin{minipage}{\textwidth}
	\centering
	\includegraphics[width= \textwidth]{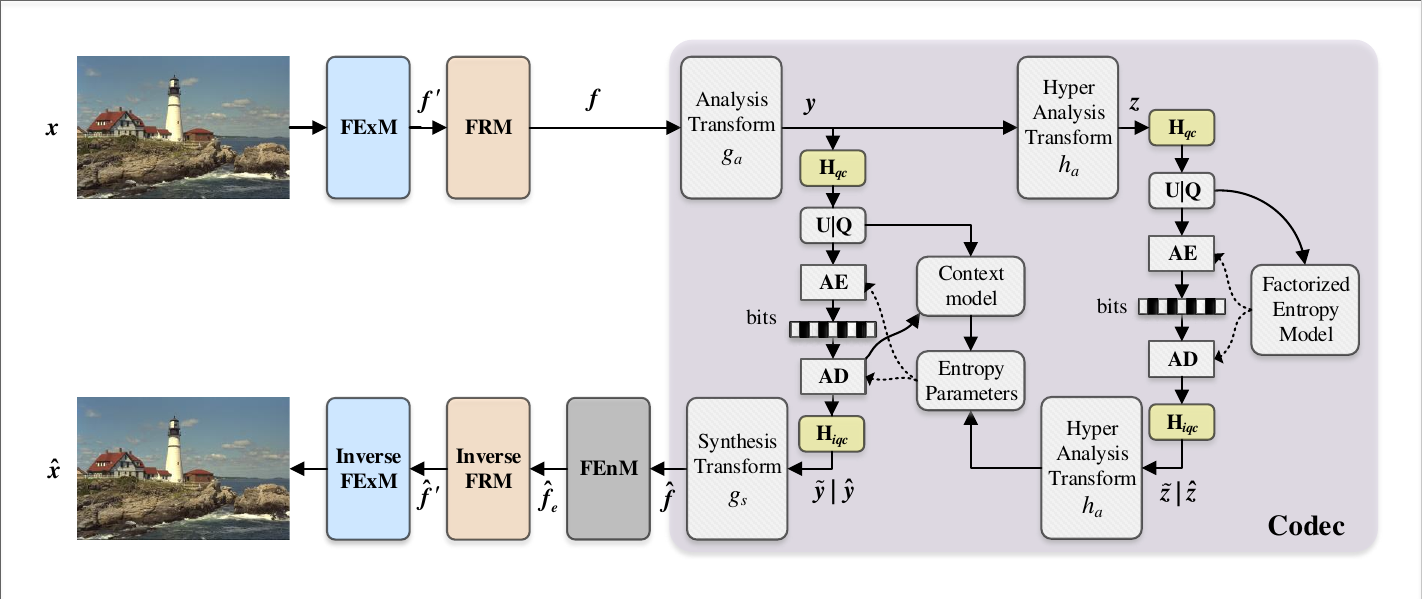}
	\caption{Overview of the proposed method. FExM denotes the feature extraction module. FRM denotes the feature refinement module. FEnM denotes the feature enhancement module. The codec adopts a joint autoregressive and hierarchical priors model, which consists of an analysis/synthesis transform, a hyper analysis/synthesis transform, a context model, an entropy parameters model, quantization and entropy coding. $\textbf{U}\mid\textbf{Q}$ denotes the quantization process, where $\textbf{U}$ represents adding uniform noise during training and $\textbf{Q}$ represents rounding quantization during testing. $\textbf{AE}$ and $\textbf{AD}$ denote arithmetic encoding and arithmetic decoding, respectively.}
	\label{fig:Overview}
	\end{minipage}
\end{figure*}
\begin{itemize}
	\item We propose a feature extraction module that shuffles the pixels of the input image and partitions the resulting image into sub-images before extracting coarse features.
 
    \item We propose a feature refinement module that uses an attention refinement block based on three-dimensional (3D) convolutions to capture feature correlations across channels, within sub-images, and across sub-images. The proposed module can effectively enhance the network's nonlinear expression ability and reduce redundancy.

	\item We propose a feature enhancement module as a post-processing step to improve the quality of decoded features. To further reduce information loss during the encoding process, we also introduce a feature enhancement loss function into the overall loss function.

	\item We propose a quantization error compensation module to deal with the mismatch between training and testing during the quantization process. The proposed module uses Fourier series to approximate the periodic changes of the quantization error and adds Laplacian noise to the quantized latent.

	\item Experimental results demonstrate the flexibility and effectiveness of the proposed modules in improving the performance of LIC methods. Our approach outperforms traditional image compression standards and state-of-the-art learning-based methods in terms of both PSNR and MS-SSIM.
\end{itemize}

The remainder of this paper is organized as follows. Section II briefly reviews the related work. Section III presents the proposed method in detail. Experimental results and conclusions are given in Section IV and V, respectively.

\section{Related work}
\subsection{Traditional image compression methods}
Image compression has been studied for decades and has resulted in a series of well-known standards, such as JPEG \cite{6}, JPEG2000 \cite{7}, and H.266/VVC \cite{8}. Traditional methods use transforms, quantization, and entropy coding. The main techniques include the discrete cosine transform (DCT), the discrete wavelet transform (DWT), context-adaptive variable-length coding (CAVLC), and context-adaptive binary arithmetic coding (CABAC). The works in \cite{8962013,9762060,8744274,9899414,8447515} have achieved significant performance gains by refining traditional methods. However, joint optimization of different techniques remains challenging with these traditional approaches.

\subsection{Learned image compression methods}
LIC methods achieve excellent performance through end-to-end joint optimization of transform networks and entropy coding.

\textbf{Transform Networks.} Toderici \textit{et al}. \cite{toderici2015variable,toderici2017full} proposed a recurrent neural network (RNN)-based image compression method that outperformed JPEG. Due to their excellent feature extraction capabilities, convolutional neural networks (CNNs) have also been widely used in image encoding \cite{theis2017lossy,balle2018variational,cheng2020learned,chen2021end,jiang23}. Ballé \textit{et al}. \cite{balle2016end} first proposed an end-to-end image compression framework using a CNN to implement nonlinear transformations. Subsequently, the work in \cite{balle2018variational} proposed a hyperprior to capture spatial dependencies in the latent representation. Minnen \textit{et al}. \cite{minnen2018joint} jointly optimized hierarchical hyperpriors and autoregressive models to achieve better RD performance. Chen \textit{et al}. \cite{chen2021end} proposed a non-local attention module (NLAM) to capture long-range correlations. Cheng \textit{et al}. \cite{cheng2020learned} later simplified the NLAM. Ma \textit{et al}. \cite{8931632} proposed iWave, a framework tailored for deriving a wavelet-like transform optimized for natural image compression. Akbari \textit{et al}. \cite{9385968} proposed a multi-resolution variable-rate image compression framework that uses generalized octave convolutions (GoConv) to factorize the feature maps into high and low resolutions. In recent years, with the explosion of self-attention, Vision Transformers (ViT) have also received widespread attention \cite{zhu2022transformer,lu2021transformer,lu2022high}. Zhu \textit{et al}. \cite{zhu2022transformer} used the Swin transformer \cite{liu2021swin} to construct nonlinear transformations. Lu \textit{et al}. \cite{lu2021transformer} used a neural transformation unit (NTU) based on a VAE architecture to extract features. Subsequently, the work in \cite{lu2022high} further adopted an integrated convolution and self-attention unit (ICSA) to achieve content-adaptive transforms.

\textbf{Entropy Models.} Building an accurate entropy model is essential for improving coding performance. Ballé \textit{et al}. \cite{balle2018variational} proposed a zero-mean Gaussian scale mixture (GSM) model for entropy coding, which achieved better performance than BPG. In the proposed GSM, the scale parameters are conditioned on a hyperprior. 
Minnen \textit{et al}. \cite{minnen2018joint} extended the work in \cite{balle2018variational} by generalizing the GSM model to a conditional Gaussian mixture model (GMM). They also combined an autoregressive model with the hyperprior. Zhou \textit{et al}. \cite{zhou2018variational} modeled the entropy model using a Laplacian distribution, which outperformed the Gaussian model. Cheng \textit{et al}. \cite{cheng2020learned} used a discrete GMM, and Fu \textit{et al}. \cite{fu2023learned} proposed a Gaussian-laplacian-logistic mixture model (GLLMM). \textcolor{black}{In addition, to enhance the encoding efficiency of the joint hyperprior and autoregressive model, some works explored parallel upper-lower modeling, such as channel-wise techniques \cite{minnen2020channel}, checkerboard context \cite{He_2021_CVPR}, and multistage context model \cite{lu2022high}. These approaches leverage parallelization to reduce the high time complexity caused by serial scanning methods.}
\begin{table}
  \caption{Notations}
  \label{tab:Notations}
  \begin{tabular}{ccl}
    \toprule
    Abbr./Symbol &Description\\
    \midrule
    FExM & Feature Extraction Module\\
	FRM & Feature Refinement Module\\
	FEnM & Feature Enhancement Module\\
	ARB & Attention Refinement Block\\
	ICRSA & Integrated Concatenated Residual and Self-Attention\\
	CRM & Concatenated Residual Module\\
	MCM  & Multi-stage Context Model\\
	DB & Dense Block\\
	QECM & Quantization Error Compensation Module\\
	U$\mid$Q & Quantization process\\
	AE/AD & Arithmetic Encoding and Arithmetic Decoding\\
	\midrule
	$\bm{S}_{i},i=1,2,...,b^2$ & Sub-images\\
	$\bm{f_x}$ & Feature corresponding to input image \bm{$x$}\\
	$\bm{f}_{\bm{S}_{i}},i=1,2,...,b^2$ & Feature corresponding to Sub-image $\bm{S}_{i}$\\
	$\bm{f'}$ & Stacked feature of $\bm{f}_{\bm{S}_1}$,$\bm{f}_{\bm{S}_2}$,$...$,$\bm{f}_{\bm{S}_{b^2}}$ and $\bm{f_x}$ \\
	$\bm{f}$ & Refined feature\\
	$\bm{\hat{f}_e}$ & Reconstructed enhanced refined feature\\
	$g_a/g_s$ & Main Encoder / Decoder\\
	$h_a/h_s$ & Hyper Encoder / Decoder\\
	$\text{H}_{qc}/\text{H}_{iqc}$ & (Inverse) Quantization Compensation Block \\
	$\bm{y}(\bm{\tilde{y}/\bm{\hat{y}}})$ & Latent (soft / hard quantized latent)\\
	$\bm{z}(\bm{\tilde{z}}/\bm{\hat{z}})$ & Hyper latent (soft / hard quantized hyper latent)\\
	$\bm{x}/ \bm{\hat{x}} $ & Input image / Reconstructed image\\
	$\bm{\psi}$ & Hyperpriors\\
    \bottomrule
	\end{tabular}
\end{table}

\begin{figure*}
	\centering
	\begin{minipage}{\textwidth}
	\centering
	\includegraphics[width=\textwidth]{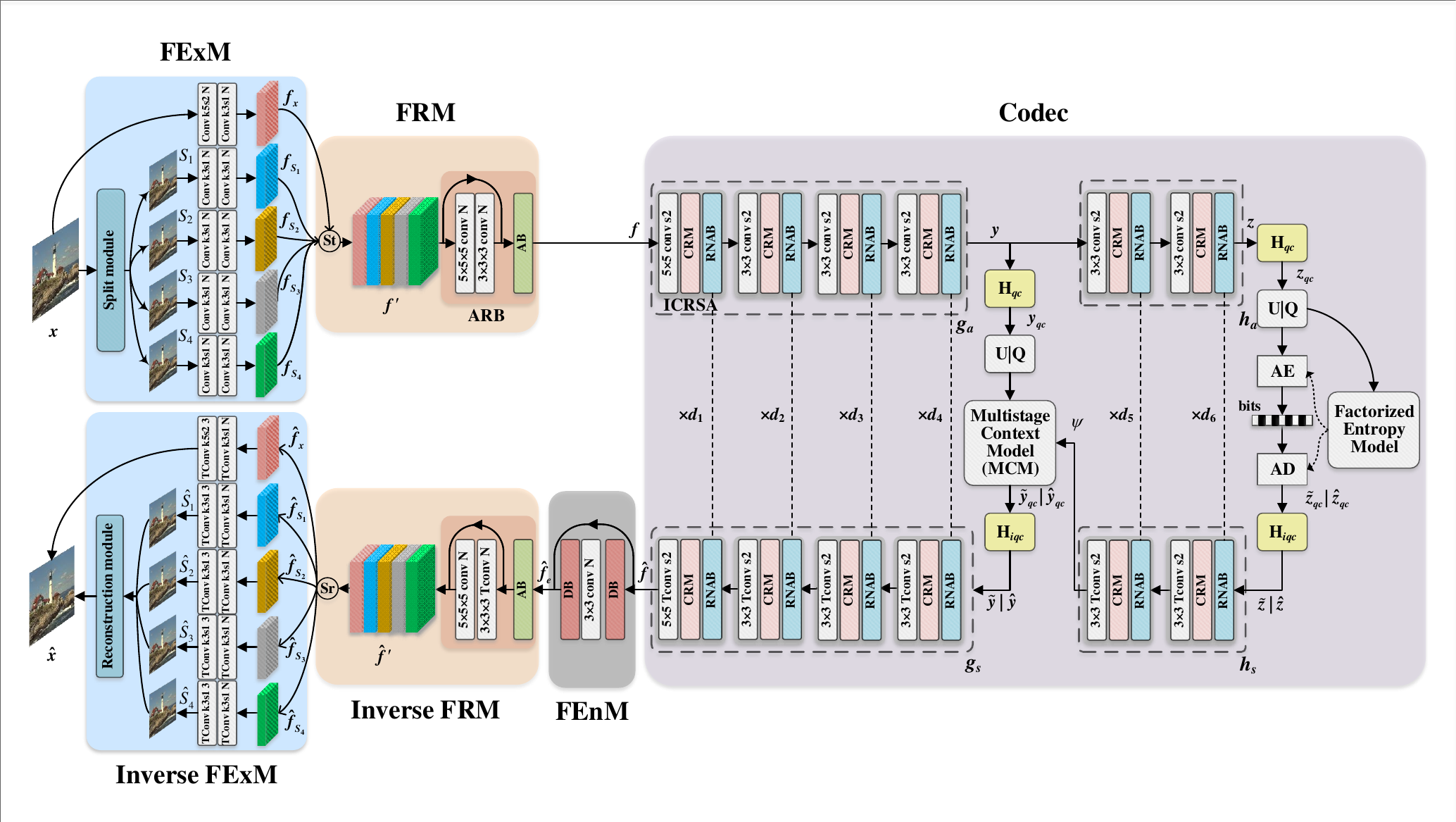}
	\caption{\textcolor{black}{Framework. The codec uses a main and a hyper encoder-decoder architecture based on VAE. FExM applies image split and feature extraction. FRM refines the stacked features with ARB, which uses an attention block (AB) to refine the features in the channel, spatial, temporal and feature dimensions. FEnM uses two dense blocks (DB) to enhance the distorted features. In the codec, each ICRSA consists of a conv layer, a concatenated residual module (CRM), and an RNAB. RNAB \cite{lu2022high} aggregates neighborhood information based on attention. $d_i(i=1,...,6)$ is the number of RNABs used at the $i$-th stage. $\mathrm{St}$, $\mathrm{Sr}$ represent stacking and re-arranging. $k5s2$ denotes a kernel with a size of $5 \times 5$ and a stride of 2. }}
	\label{fig:framework}
	\end{minipage}
\end{figure*}

\textbf{Quantization.} One of the primary challenges is maintaining a differentiable and continuous process for backpropagation-based training, which contrasts with the non-differentiable nature of hard quantization operations. To tackle this issue, non-differentiable quantization is approximated using differentiable techniques. This approximation can be achieved by introducing additive uniform noise~\cite{balle2016end,balle2018variational,minnen2018joint,fu2023learned,kim2022joint}, straight-through estimator (STE)~\cite{bengio2013estimating}, and soft-to-hard annealing~\cite{agustsson2017soft,yang2020improving}. Most prevalent approaches opt for soft quantization with additive uniform noise as a replacement for hard quantization during training. However, this approach can lead to a mismatch between training and testing. Eirikur \textit{et al}.~\cite{agustsson2020universally} introduced universal quantization and soft rounding methods to bridge the gap between training and testing. Guo \textit{et al}.~\cite{guo2021soft} proposed a soft-then-hard strategy to learn latent representations and address the aforementioned mismatch issue. Fu \textit{et al}.~\cite{fu2022asymmetric} introduced a post-quantization filter to minimize the difference between the latent and dequantized latent representations. However, these methods often overlook the characteristics and statistical distribution of quantization errors. To alleviate the mismatch problem, we propose a quantization error compensation method based on Fourier series approximation and Laplacian noise. Our paper extends our previous work~\cite{jiang23} by adding feature extraction, refinement, and enhacement modules.

\section{Proposed method}
\subsection{Overview}

We propose four modules to enhance the performance of state-of-the-art LIC methods: a feature extraction module (FExM), a feature refinement module (FRM), a feature enhancement module (FEnM), and a quantization error compensation module (QECM), as shown in Figure \ref{fig:Overview}. Most LIC methods are based on a VAE architecture, consisting of a main encoder-decoder and a hyper encoder-decoder. In this work, we adopt the network architecture described in \cite{lu2022high} as our backbone LIC network. Our coding framework is illustrated in Figure \ref{fig:framework}. For clarity, we summarize the notations used in our approach in Table \ref{tab:Notations}.

\subsection{Feature Extraction Module}
We propose a feature extraction module (FExM) that splits the image into sub-images to better exploit correlations within the image. 

\begin{figure}[htbp]
	\centering
	\includegraphics[width=\linewidth]{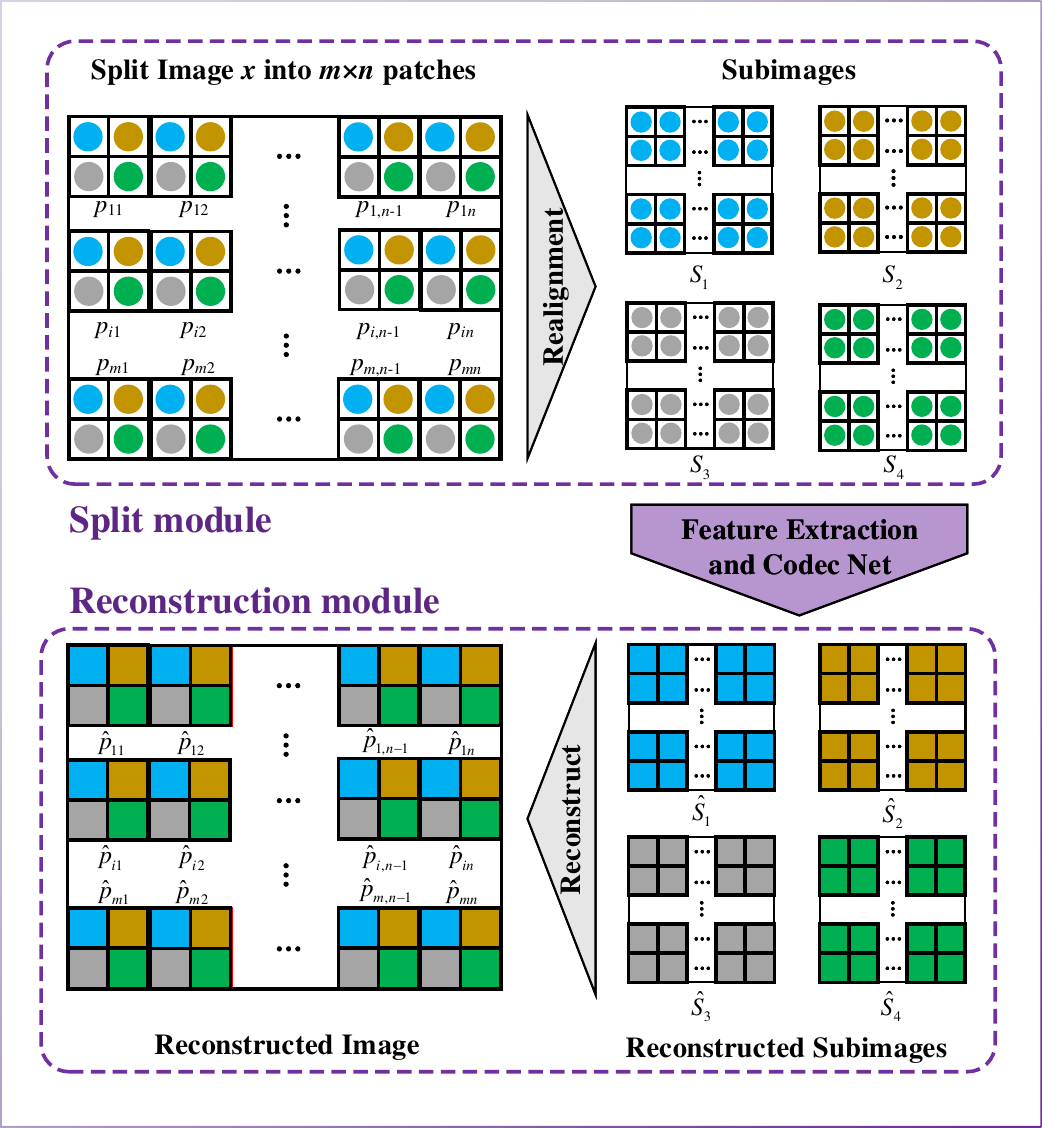}
	\caption{Split module and reconstruction module. Solid circles represent the original pixels, while solid rectangles represent the reconstructed pixels.}
	\label{fig:split}
\end{figure}

In high-resolution images, inherent spatial redundancy is heightened. Processing entire images imposes resource burdens and poses challenges in effectively extracting global and local features. In video sequences, the content of adjacent frames may exhibit a high degree of similarity, allowing learning-based video coding methods to leverage both inter-frame and intra-frame correlation to remove redundancy. Inspired by this, we convert an image into multiple sub-images (frames) and remove both intra and inter-frame redundancy. First, we split the input image $\bm{x} \in \mathbb{R}^{H \times W \times 3}$ into $m \times n$ patches, where each patch consists of $b \times b$ pixels. We denote the patch located in the $i$-th row and $j$-th column, $i = 1,...,m$, $j = 1,...,n$, by $\bm{p}_{ij} \in \mathbb{R}^{b \times b \times 3}$. Then, we extract the pixels at the corresponding positions in each patch and realign them according to their indices $i$ and $j$. As a result, we obtain $b^2$ sub-images $\bm{S}_{t} \in \mathbb{R}^{W/b \times H/b \times 3}, t=1,2,...,b^2$. Figure \ref{fig:split} illustrates the case $b=2$. Through the splitting approach, the sub-images $\bm{S}_1$,$\bm{S}_2$,$...$,$\bm{S}_{b^2}$ are highly similar to each other and can be considered as $b^2$ frames in a video with temporal correlations. Furthermore, $\bm{S}_1$,$\bm{S}_2$,$...$,$\bm{S}_{b^2}$ retain all pixel information of the input \bm{$x$}. Our approach has similarity with ViTs \cite{dosovitskiy2021an}, which segment images into patches to capture both local and global features, using patch merging to increase the number of channels while simultaneously reducing the resolution. However, unlike ViTs, our approach exploits inter-frame correlations. Increasing the patch size increases the number of low-resolution frames. However, increasing inter-frame correlation is achieved at the cost of reduced intra-frame correlation. The effect of varying patch size on performance is evaluated in Section IV. E.

To extract features, we use cascaded convolutional layers. Specifically, we apply two convolutional layers separately to $\bm{S}_t$ to obtain the corresponding features $\bm{f}_{\bm{S}_t}$, $t=1,2,...,b^2$. Moreover, we use a convolutional layer with a stride of 2 to extract the feature of \bm{$x$}, denoted as \bm{$f_{x}$}, as shown in Figure \ref{fig:framework}.

The inverse FExM follows the reverse process of the FExM and adopts the same network architecture using transposed convolutions. We obtain the final reconstructed image \bm{$\hat{x}$} from the reconstructed patches $\bm{\hat{p}}_{ij}$, $i = 1,...,m$, $j = 1,...,n$, as shown in Figure \ref{fig:split}.

\subsection{Feature Refinement Module}
To exploit the correlation between features, we propose a feature refinement module (FRM) that includes feature stacking and an attention refinement block.

\textbf{Feature stacking.} After feature extraction in FExM, we obtain $b^2+1$ three-dimensional features, $\bm{f}_{\bm{S}_1}$,$\bm{f}_{\bm{S}_2}$,$...$,$\bm{f}_{\bm{S}_{b^2}}$ and $\bm{f_x}$ $\in \mathbb{R}^{W/b \times H/b \times N}$. The highly similar sub-images $\bm{S}_1$,$\bm{S}_2$,$...$,$\bm{S}_{b^2}$, not only exhibit intra redundancy but also inter redundancy. To capture the correlation between features, we construct a new feature by stacking the features $\bm{f}_{\bm{S}_1}$,$\bm{f}_{\bm{S}_2}$,$...$,$\bm{f}_{\bm{S}_{b^2}}$ and $\bm{f_x}$ in an additional temporal dimension. Moreover, we consider \bm{$f_{x}$} to facilitate feature learning. In this way, we obtain a feature $\bm{f^{'}}\in \mathbb{R}^{(b^2+1) \times W/b \times H/b \times N}$ such that 
\begin{equation}
  \bm{f^{'}} = \bm{\mathcal{F}}(\bm{f}_{\bm{S}_1},\bm{f}_{\bm{S}_2},...,\bm{f}_{\bm{S}_{b^2}},\bm{f_x}),  
\end{equation}
where $\bm{\mathcal{F}}$ is the feature stacking operator.

\begin{figure}
	\centering
	\includegraphics[width=\linewidth]{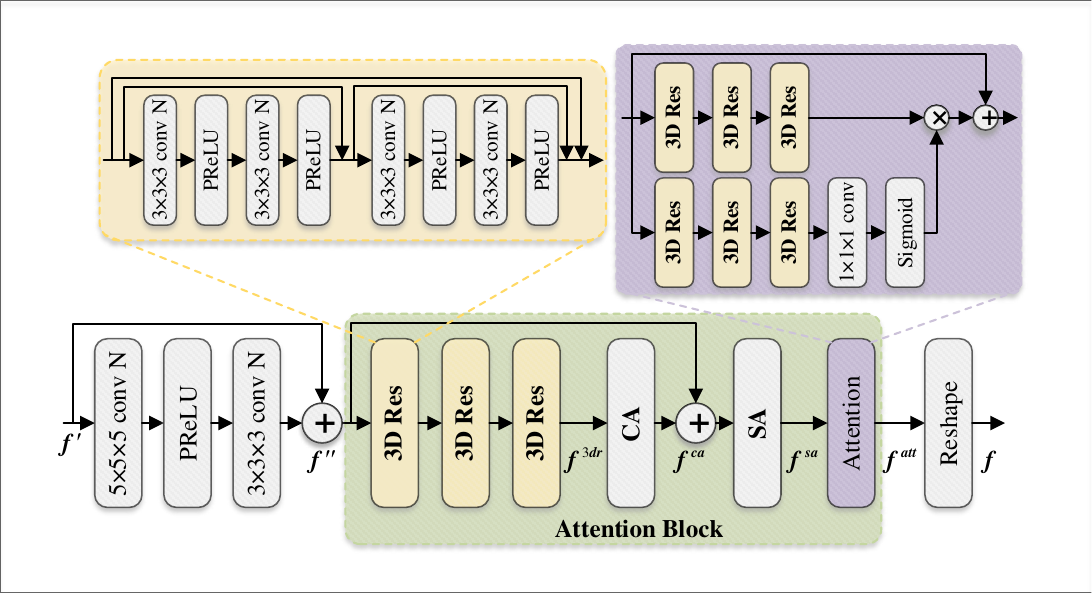}
	\caption{Attention Refinement Block. The 3D Res-b structure (as shown in Figure \ref{fig:3D Res}(b)) is used as an example.}
	\label{fig:ARB}
\end{figure}

\textbf{Attention Refinement Block.} To extract features more effectively, we add an attention refinement block (ARB), as indicated in Figure \ref{fig:ARB}. A previous work \cite{duan2022end} also proposed a feature attention block. However, unlike this work, our attention refinement block uses 3D convolutions for feature extraction. First, we use two 3D convolutional layers to extract features from $\bm{f^{'}}$ and use the shortcut structure. These features are fed into an attention block (AB). The attention block consists of three 3D convolution residual blocks (3D Res), a channel attention (CA) block \cite{zhang2018image} with global residual, a spatial attention (SA) block \cite{woo2018cbam}, and a feature attention block \cite{cheng2020learned}, where we reconstruct the attention model using 3D convolutions. CA and SA focus more on the feature information in channels and space, respectively, and the attention uses cascaded 3D Res to pay attention to more challenging feature information. The output is reshaped to 3D space to obtain $\bm{f}$.

\begin{figure}
	\centering
	\includegraphics[width=\linewidth]{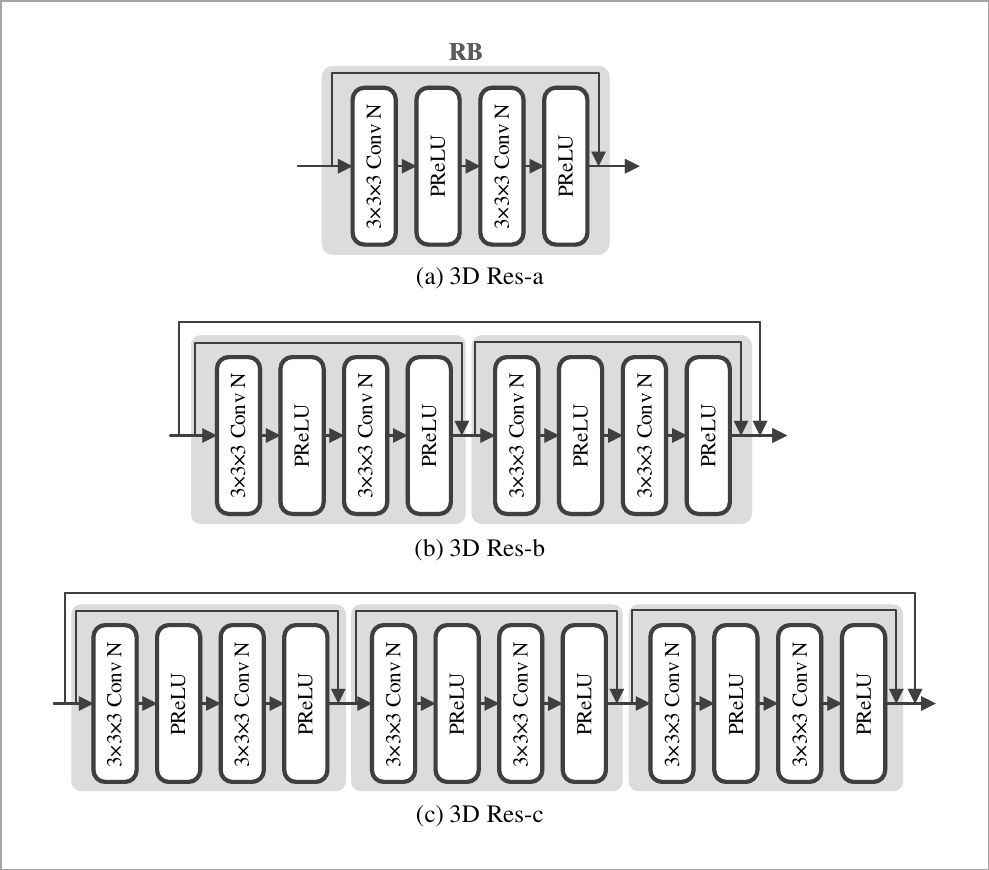}
	\caption{3D Res blocks. (a) 3D Res-a: Basic 3D Residual block (RB), (b) 3D Res-b: Two-level concatenated 3D residual block,(c) 3D Res-c: Three-level 3D residual block}
	\label{fig:3D Res}
\end{figure}

A concatenated residual block can effectively enhance the encoding performance \cite{fu2023learned}. To further eliminate feature correlation, we also design two deeper 3D convolution residual blocks, as shown in Figure \ref{fig:3D Res}. Figure \ref{fig:3D Res} (a) displays the basic 3D Res. Unlike the method in \cite{fu2023learned}, we reconstruct the residual block (RB) using 3D convolutions and use PReLU as the activation function. Figure \ref{fig:3D Res} (b) and (c) display the two-level and three-level cascaded residual blocks, respectively, with added shortcut connections. In this way, concatenated 3D Res blocks can obtain a larger receptive field, which is helpful for feature extraction and correlation removal. In section IV, we compare the performance of different 3D Res blocks.

\subsection{Codec}
The codec adopts the main and hyper encoder-decoder structure of VAE \cite{balle2018variational}, which can be formulated as:
\begin{equation}
  \bm{y} = g_a(\bm{x};\bm{\theta_{me}}),
\end{equation}
\begin{equation}
  \bm{\tilde{y}}=\boldsymbol{y}+\Delta \boldsymbol{y}, \Delta \boldsymbol{y} \sim \mathrm{U}(-0.5,0.5),
\end{equation}
\begin{equation}
  \bm{\hat{y}} = \textbf{Q}(\bm{y}) = \operatorname{round}(\boldsymbol{y}),
\end{equation}
\begin{equation}
  \bm{\hat{x}} = g_s(\bm{\hat{y}};\bm{\theta_{md}}),
\end{equation}
where \bm{$x$} is fed into the main encoder $g_a$ to generate \bm{$y$}, \bm{$\tilde{y}$} is the soft quantized latent feature with additive uniform noise $\Delta \boldsymbol{y}$ for training, \bm{$\hat{y}$} is the hard quantized latent feature obtained by rounding for testing, 
\bm{$\hat{x}$} is obtained by the main decoder $g_s$, \bm{$\theta_{me}$} and \bm{$\theta_{md}$} denote the optimized parameters of $g_a$ and $g_s$. A hyperprior network and a context model are used to effectively improve the entropy model.

We adopt Tiny-LIC \cite{lu2022high} as the backbone. Tiny-LIC uses an integrated convolution and self-attention unit (ICSA) and a multi-stage context model (MCM) to improve encoding efficiency through parallel processing while ensuring encoding performance. MCM can be flexibly replaced with other parallelized methods. We propose an integrated concatenated residual and self-attention unit (ICRSA) by introducing a two-stage concatenated residual module (CRM) \cite{fu2023learned} between the conv and the residual neighborhood attention block (RNAB) within each ICSA, as shown in Figure \ref{fig:framework}. The two-stage CRM represents the cascading of two residual blocks connected by a shortcut between the input and output. This configuration allows for a larger receptive field and the removal of more spatial redundancy.
\begin{figure}
	\centering
	\includegraphics[width=1\linewidth]{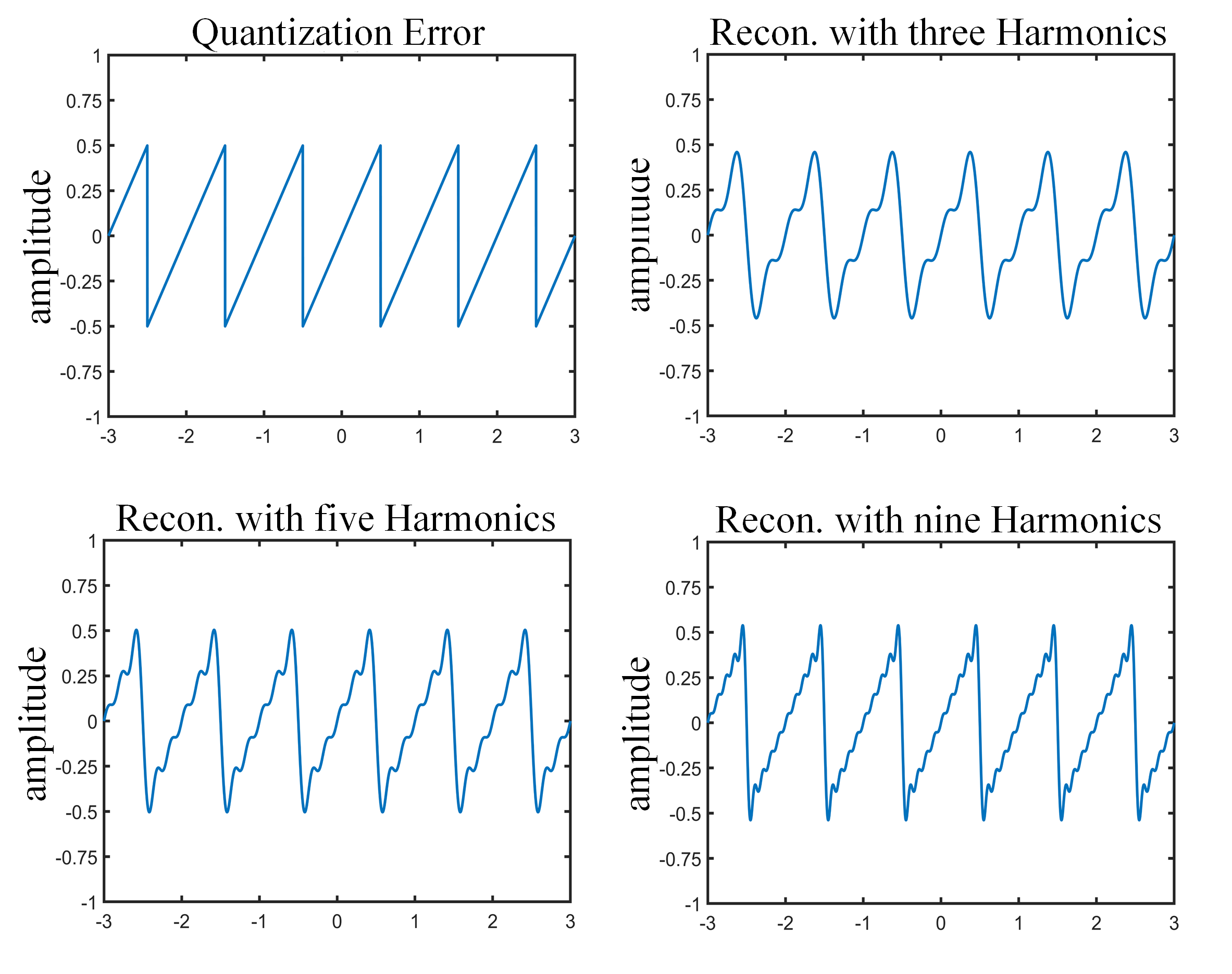}
	\caption{Quantization error approximation with increasing number of harmonics. Recon. represents reconstruction.}
	\label{fig:wave-icip}
\end{figure}
\subsection{Quantization Error Compensation Module}
Quantization is a crucial step in lossy image compression, but it unavoidably results in information loss. The quantization error $Q_{e r r}$ is the difference between the initial input $\boldsymbol{y}$ and the quantized output $\bm{\hat{y}}$. That is, 
\begin{eqnarray}
Q_{e r r}(\boldsymbol{y})=\boldsymbol{y}-\bm{\hat{y}}.
\label{eq:quan_error}
\end{eqnarray}

This error has the shape of a sawtooth wave, as illustrated in Figure \ref{fig:wave-icip}. While quantization is not differentiable, the quantization error is influenced by a cyclic variation in signal distribution. This cyclic behavior can be approximated as
\begin{eqnarray}
Q_{e r r}(\boldsymbol{y}) = s(\boldsymbol{y}),
\end{eqnarray}
where $s(\boldsymbol{y})$ is the sawtooth wave with period 1 and equal to ${1}/{2\bm{y}}$ over the interval $[-1/2,1/2]$. The Fourier series of this function is
\begin{eqnarray}
\begin{aligned}
s(\boldsymbol{y}) & =\frac{a_0}{2}+\sum_{\mathrm{n}=1}^{\infty}\left[a_{\mathrm{n}} \cos (n \boldsymbol{y})+b_{\mathrm{n}} \sin (n \boldsymbol{y})\right], \\
& =\frac{1}{\pi} \sum_{n=1}^{\infty} \frac{(-1)^{n+1}}{n} \sin (n \boldsymbol{y}),
\end{aligned}
\end{eqnarray}

Consequently, Eq. \ref{eq:quan_error} can be rewritten as:
\begin{eqnarray}
\bm{\hat{y}}&=&\boldsymbol{y} - Q_{e r r}(\boldsymbol{y}),\nonumber\\
&=&\boldsymbol{y}-\frac{1}{\pi} \sum_{n=1}^{\infty} \frac{(-1)^{n+1}}{n} \sin (n \boldsymbol{y}),
\label{eq:quan}
\end{eqnarray}
where $n$ denotes the index of the harmonic. The effects of the harmonic approximation are illustrated in Figure \ref{fig:wave-icip}. We can select various values for $n$ to regulate the level of approximation to the original quantization error. Note that using a finite sum of terms in the Fourier series approximation for the quantization error function guarantees differentiability.

During the training process, quantization is approximated by introducing additive uniform noise. To take this approximation into account, we include a quantization error compensation module, represented as the second component in Eq. \ref{eq:quan}, into the baseline model, as illustrated in Figure \ref{fig:framework}. For the latent $\boldsymbol{y}$ and hyper latent $\boldsymbol{z}$, a quantization compensation block $h_{qc}$ is added at the encoder side, and an inverse quantization compensation block $h_{iqc}$  is added at the decoder side. The compensation module $h_{qc}$ takes the inputs $\boldsymbol{y}$ and $\boldsymbol{z}$ and outputs $\boldsymbol{y}_{\boldsymbol{qc}}$ and $\boldsymbol{z}_{\boldsymbol{q c}}$, where
\begin{eqnarray}
\boldsymbol{y}_{\boldsymbol{qc}}&=&\boldsymbol{y}-s(\boldsymbol{y}),\\
\boldsymbol{z}_{\boldsymbol{qc}}&=&\boldsymbol{z}-s(\boldsymbol{z}).
\end{eqnarray}

Correspondingly, $h_{iqc}$ takes the inputs $\tilde{\boldsymbol{y}}_{\boldsymbol{qc}}$ and $\tilde{\boldsymbol{z}}_{\boldsymbol{qc}}$ and outputs $\tilde{\boldsymbol{y}}$ and $\tilde{\boldsymbol{z}}$, where
\begin{eqnarray}
\tilde{\boldsymbol{y}}&=&\tilde{\boldsymbol{y}}_{\boldsymbol{qc}}+s(\tilde{\boldsymbol{y}}_{\boldsymbol{qc}}),\\
\tilde{\boldsymbol{z}}&=&\tilde{\boldsymbol{z}}_{\boldsymbol{qc}}+s(\tilde{\boldsymbol{z}}_{\boldsymbol{qc}}).
\end{eqnarray}

As noted in~\cite{jiang23}, the quantization error follows a Laplace distribution instead of a Gaussian distribution in LIC methods. Consequently, we add Laplacian random noise to the quantized value during testing. This further helps mitigate the discrepancy between training and testing.This gives 
\begin{eqnarray}
\hat{\boldsymbol{y}}&=&\hat{\boldsymbol{y}}_{\boldsymbol{q c}}+s\left(\hat{\boldsymbol{y}}_{\boldsymbol{q c}}+\Delta \boldsymbol{n}\right), \Delta \boldsymbol{n} \sim \mathrm{L}(\mu, b),\\
\hat{\boldsymbol{z}}&=&\hat{\boldsymbol{z}}_{\boldsymbol{q c}}+s\left(\hat{\boldsymbol{z}}_{\boldsymbol{q c}}+\Delta \boldsymbol{n}\right), \Delta \boldsymbol{n} \sim \mathrm{L}(\mu, b),
\end{eqnarray}
where $\Delta \boldsymbol{n}$ denotes additive noise following a Laplace distribution $\mathrm{L}(\mu, b)$ limited to $(-1/2, 1/2)$, with $\mu$ and $b$ representing the mean and scale, respectively.

\subsection{Feature Enhancement Module}

Quantization is commonly used in image and video compression to reduce bitrate, although it introduces errors. When the feature undergoes encoding and decoding by a codec, the lossy quantization inevitably results in information loss. Consequently, we introduce a feature enhancement module (FEnM) to improve the quality of decoded features and alleviate information loss, as depicted in Figure 2. The FEnM module comprises two dense blocks (DBs) and a convolutional layer, interconnected through shortcuts.

The following feature enhancement loss function is incorporated into the overall cost function:
\begin{equation}
  \mathcal{L}_{FE} =  \| \bm{\hat{f}_{e}} - \bm{f} \|^2,
\label{eq:FEnM loss}
\end{equation}
where $\bm{\hat{f}_{e}}$ represents the output of the FEnM, $\bm{f}$ represents the refined feature after FRM.

\subsection{Loss function}
LIC can be thought of as an RD optimization problem based on Lagrange multipliers. To improve the performance, the FEnM loss defined in Eq. \ref{eq:FEnM loss} is added to the overall cost function as follows:
\begin{equation}
  \begin{aligned} \mathcal{L}= & \mathcal{R}(\hat{\boldsymbol{y}})+\mathcal{R}(\hat{\boldsymbol{z}})+\lambda \cdot \mathcal{D}(\boldsymbol{x}, \hat{\boldsymbol{x}}) + \lambda_{e} \cdot \mathcal{L}_{FE} \\ = & \mathbb{E}\left[-\log _2\left(p_{\hat{\boldsymbol{y}} \mid \hat{\boldsymbol{z}}}(\hat{\boldsymbol{y}} \mid \hat{\boldsymbol{z}})\right)\right]+\mathbb{E}\left[-\log _2\left(p_{\hat{\boldsymbol{z}}}(\hat{\boldsymbol{z}})\right)\right] \\ & +\lambda \cdot \mathcal{D}(\boldsymbol{x}, \hat{\boldsymbol{x}}) + \lambda_{e} \cdot \| \bm{\hat{f}_{e}} - \bm{f} \|^2,
  \end{aligned}
\end{equation}
where rate $\mathcal{R}$ is composed of the entropy of the latent representation $\mathcal{R}(\hat{\boldsymbol{y}})$ and the entropy of the hyper latent representation $ \mathcal{R}(\hat{\boldsymbol{z}})$, $\mathcal{D}(\bm{x}, \hat{\bm{x}})$ denotes the reconstruction error between $\boldsymbol{x}$ and $\hat{\boldsymbol{x}}$, $\lambda$ and $\lambda_{e}$ are used to control the trade-off between rate and distortion.

\section{Experimental results and analysis}
\subsection{Experimental Setup}

\textbf{Training.} We used the ImageNet2017 dataset \cite{deng2009imagenet} as the training set. We randomly cropped all images into patches of size $256 \times 256$, resulting in over 1M patches in total. We trained the model for 600 epochs on Nvidia RTX3090Ti. We set the learning rate to 1e-4 in the first 400 epochs, to 1e-5 in the following 150 epochs, and to 1e-6 in the remaining epochs. 

\textbf{Settings.} We implemented our model with the compressAI Pytorch library \cite{begaint2020compressai}, using the Adam optimizer with a batch size of 8. The number N of convolution channels was set to 128. We set the model parameters of Tiny-LIC as in \cite{lu2022high}. In the main encoder-decoder, we set the numbers of RNABs, i.e., $d_1,d_2,d_3$ and $d_4$ to 2, 2, 6 and 2, the hyper encoder-decoder parameters, $d_5$ and $d_6$ to 2, and $\lambda$ to 0.0018, 0.0035, 0.0067, 0.013, 0.025, 0.0483 to minimize the mean squared error (MSE). In the initial 200 epochs, $\lambda_{e}$ was set to 1 to facilitate the learning of a reconstructed enhanced refined feature $\bm{\hat{f}_e}$ that closely approximates the refined feature $\bm{f}$, with the objective of improving reconstruction quality. Subsequently, for the remaining epochs, $\lambda_{e}$ was set to 0, enabling end-to-end training of the entire network guided by RD optimization, with the goal of achieving optimal compression performance. The number of harmonics in the Fourier series was set to 5.

\textbf{Testing.} We evaluated the proposed method on two commonly used datasets: the Kodak dataset \cite{franzen1999kodak} containing 24 images with resolution 768x512 and the CLIC dataset \cite{clic2020} containing 41 images with higher resolutions.
\begin{table}
  \begin{center}
	  \caption{Average BD-rate(\%) improvement for the Kodak and CLIC datasets. BPG is used as the anchor. Bold font indicates the best performance.}
	  \label{tab:bd-rate}
	  \begin{tabular}{ccc}
	    \toprule
	    \textbf{Methods} & \textbf{Kodak} & \textbf{CLIC} \\
	    \midrule
		Ballé \cite{balle2018variational} &  4.3  & 1.2\\
		Minnen \cite{minnen2018joint} &  -11.9  & -18.6\\
		Cheng \cite{cheng2020learned} &  -19.0  & -24.9\\
		H.266/VVC \cite{VVC} & - 20.9  & -25.9\\
		Xie \cite{xie2021enhanced} & - 21.8  & -28.6\\
		Duan \cite{Duan_2023_WACV} & - 22.3  & -29.4\\
		Tiny-LIC \cite{lu2022high}  &  - 22.9  & -29.9\\
		STF \cite{Zou_2022_CVPR} & - 23.9  & -30.2\\
		LIC-TCM \cite{liu2023learned}  &  - 24.7  & -31.9\\
		Proposed-Tiny  &  - 23.5  & -31.1 \\
		\textbf{Proposed}  &  \textbf{- 25.4}  & \textbf{-33.4}\\
	  \bottomrule
	\end{tabular}
  \end{center}
\end{table}
\begin{table}
  \begin{center}
	\caption{Number of network parameters and time complexity of LIC methods on the Kodak dataset.}
	\label{tab:complexity}
	\resizebox{1.0\linewidth}{!}{
	\begin{tabular}{cccc}
	\toprule
	\textbf{Method} & \textbf{Parameters} & \textbf{Enc. Time} & \textbf{Dec. Time} \\
	\midrule
	Minnen \cite{minnen2018joint} & 25.5M & 4.33s & 9.18s \\
	Cheng \cite{cheng2020learned}  & 29.6M & 5.89s & 11.49s  \\
	Xie \cite{xie2021enhanced}  & 50.0M & 5.46s & 12.96s  \\
	Tiny-LIC \cite{lu2022high} & 28.3M & 0.17s & 0.12s  \\
	STF \cite{Zou_2022_CVPR}  & 99.9M & 0.23s & 0.25s  \\
	LIC-TCM \cite{liu2023learned}  & 44.9M & 0.30s & 0.32s  \\
	Proposed-Tiny  & 33.5M & 0.20s & 0.15s \\
	Proposed  & 49.5M & 0.31s & 0.35s \\
	\bottomrule
	\end{tabular}
	}
  \end{center}	
	Testing environment: Intel Xeon(R) Silver 4210 CPU, NVIDIA RTX 3090Ti GPU, Ubuntu 20.04.
\end{table}

\begin{table}
  \begin{center}
	\caption{Computational complexity of the proposed modules.}
	\label{tab:complexity_module}
	\resizebox{1.0\linewidth}{!}{
	\begin{tabular}{ccccc|c}
	\toprule
	\multicolumn{5}{c|}{ Proposed modules } & \multirow{2}{*}{ GMACs } \\
	\cmidrule{1-5}
	\textcolor{black}{Baseline} & FExM & \textcolor{black}{FRM} & \textcolor{black}{FEnM} & \textcolor{black}{QECM} & \\
	\midrule
	 \textcolor{black}{$\checkmark$} &  &  &  &  & \textcolor{black}{192G} \\ 
	 \textcolor{black}{$\checkmark$} & \textcolor{black}{$\checkmark$} &  & &  & \textcolor{black}{236G} \\ 
  \textcolor{black}{$\checkmark$} & \textcolor{black}{$\checkmark$} & \textcolor{black}{$\checkmark$} & &  &  \textcolor{black}{347G}\\ 
	 \textcolor{black}{$\checkmark$} & \textcolor{black}{$\checkmark$} & \textcolor{black}{$\checkmark$} & \textcolor{black}{$\checkmark$} &   & \textcolor{black}{355G} \\
	 \textcolor{black}{$\checkmark$} & \textcolor{black}{$\checkmark$} & \textcolor{black}{$\checkmark$} & \textcolor{black}{$\checkmark$} & \textcolor{black}{$\checkmark$} & \textcolor{black}{356G} \\
	\bottomrule
	\end{tabular}
	}
  \end{center}	
\end{table}

\subsection{RD Performance}
To assess the RD performance of the proposed method, we compared it with the following learning-based image coding methods: Ballé \cite{balle2018variational}, Minnen \cite{minnen2018joint}, Cheng \cite{cheng2020learned}, Xie \cite{xie2021enhanced}, Duan \cite{Duan_2023_WACV}, STF \cite{Zou_2022_CVPR}, Tiny-LIC \cite{lu2022high} and  LIC-TCM \cite{liu2023learned}. We obtained the source codes of these methods from their official GitHub repositories and retrained the models in the same way as our model. In addition, we compared our method with traditional codecs such as JPEG \cite{6}, BPG \cite{bellard2017bpg} and H.266/VVC intra codec \cite{VVC} (VTM 18.0), and evaluated them using the CompressAI platform \cite{begaint2020compressai}. The proposed method was integrated in Tiny-LIC. \textcolor{black}{As the proposed method increases the complexity of Tiny-LIC, we also propose a lightweight version, namely ``Proposed-Tiny". Specifically, in ``Proposed-Tiny", FExM uses a single $3 \times 3$ convolutional layer, ARB is simplified to a single $5 \times 5 \times 5$ convolutional layer, AB uses only one 3D Res block, FEnM uses only one DB, and CRM is replaced by a $3 \times 3$ convolution.}
\begin{figure*}
	\centering
	\begin{minipage}[b]{0.48\linewidth}
	  \includegraphics[width=\linewidth]{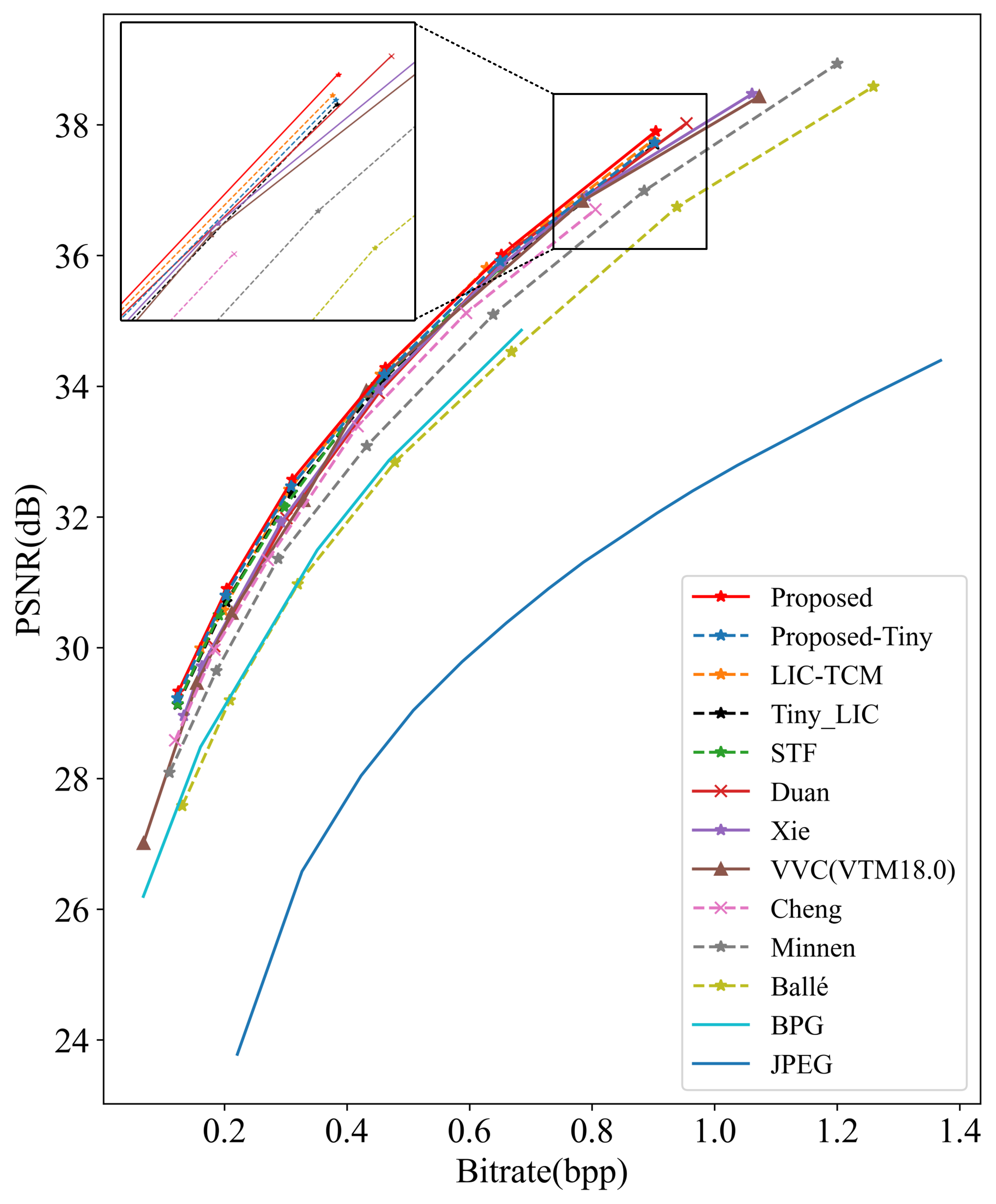}
	  \centerline{\textcolor{black}{(a) RD performance on Kodak.}}\medskip
	\end{minipage}
	\begin{minipage}[b]{0.48\linewidth}
	  \includegraphics[width=\linewidth]{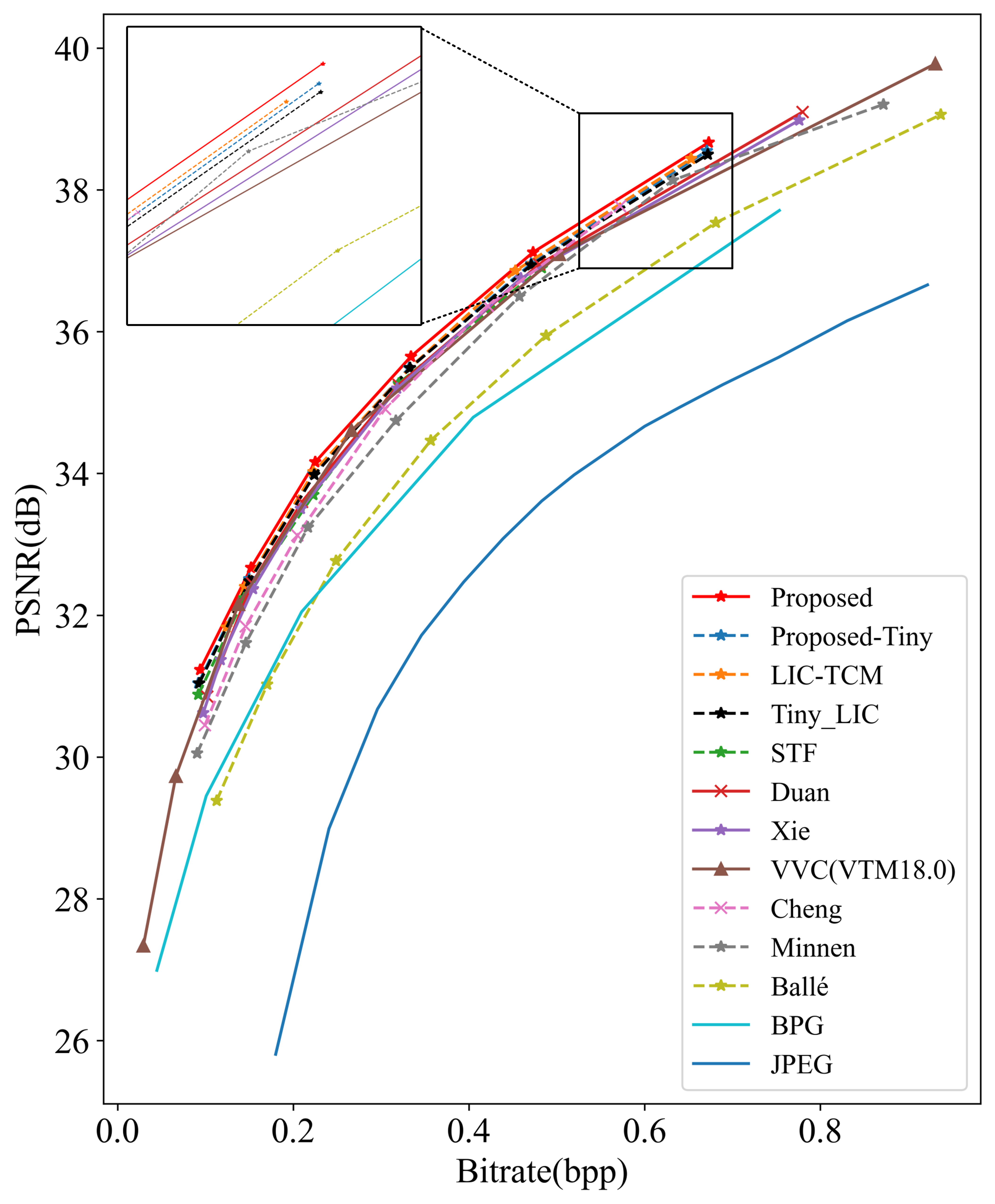}
	  \centerline{\textcolor{black}{(b) RD performance on CLIC.}}\medskip
	\end{minipage}
	\centering
	\caption{\textcolor{black}{Performance evaluation on the Kodak and CLIC datasets. VTM is the VVC test model. The bitrate is expressed in bits per pixels (bpp).}}
	\label{fig:mse_rd_curve}
\end{figure*}

\begin{table*}
  \begin{center}
	\caption{Ablation study. Baseline is the method using the method in \cite{lu2022high} as the backbone. The bitrate is in bpp. The best performance is highlighted in bold.}
	\label{tab:ablation}
	\begin{tabular}{l|ccc|ccc|c}
	\toprule
	\textcolor{black}{\textbf{Method}} & \textbf{$\lambda_1$} & \textbf{PSNR $\uparrow$} & \textbf{Bitrate  $\downarrow$} & \textbf{$\lambda_2$} & \textbf{PSNR $\uparrow$} & \textbf{Bitrate  $\downarrow$} & \textcolor{black}{\textbf{BD-rate $\downarrow$}} \\
	\midrule
	\textcolor{black}{Baseline} & 0.0035 & 30.72 & 0.206 & 0.0483 & 37.76 & 0.910 & \textcolor{black}{0}\\ 
	\textcolor{black}{Baseline+FExM+FRM} & 0.0035 & 30.80 & 0.208 & 0.0483 & 37.86 & 0.913 & \textcolor{black}{-1.1}\\ 
	\textcolor{black}{Baseline+FExM+FRM+FEnM (w/o $\mathcal{L}_{FE}$)} & 0.0035 & 30.83 & 0.204 & 0.0483 & 37.91 & 0.909 & \textcolor{black}{-1.9}\\ 
	\textcolor{black}{Baseline+FExM+FRM+FEnM} & 0.0035 & 30.89 & 0.204 & 0.0483 & 37.99 & 0.909 & \textcolor{black}{-2.5}\\ 
	\textcolor{black}{\textbf{Baseline+FExM+FRM+FEnM+QECM}} & \textbf{0.0035} & \textbf{30.95} & \textbf{0.203} & \textbf{0.0483} & \textbf{38.08} & \textbf{0.905} & \textcolor{black}{-3.6}\\ 
	\bottomrule
	\end{tabular}
  \end{center}
\end{table*}

Figure \ref{fig:mse_rd_curve} (a) and (b), respectively, show the RD curves of the methods on the Kodak and the CLIC datasets. In our method, $b$ was set to 2, i.e., the patch size was $2 \times 2$. The proposed method achieved the best RD performance. Table \ref{tab:bd-rate}, in which BPG is used as the anchor, shows the average Bjontegaard delta (BD-rate) \cite{bjontegaard2001calculation} reductions on the test sets. When tested on the Kodak dataset, the proposed method achieved a 2.5\% bitrate reduction over Tiny-LIC, and a 4.5\% reduction over H.266/VVC. When tested on the CLIC dataset, the proposed method achieved a 3.5\% bitrate reduction over Tiny-LIC, and a 7.5\% reduction over H.266/VVC. As the image resolution in CLIC is higher than in Kodak, the correlation between sub-images is stronger. This enhanced the ability to extract both intra and inter-correlations, resulting in better RD performance. 
\begin{figure*}[htbp]
	\centering
	\begin{minipage}{\textwidth}
	\centering
	\includegraphics[width=0.99\textwidth]{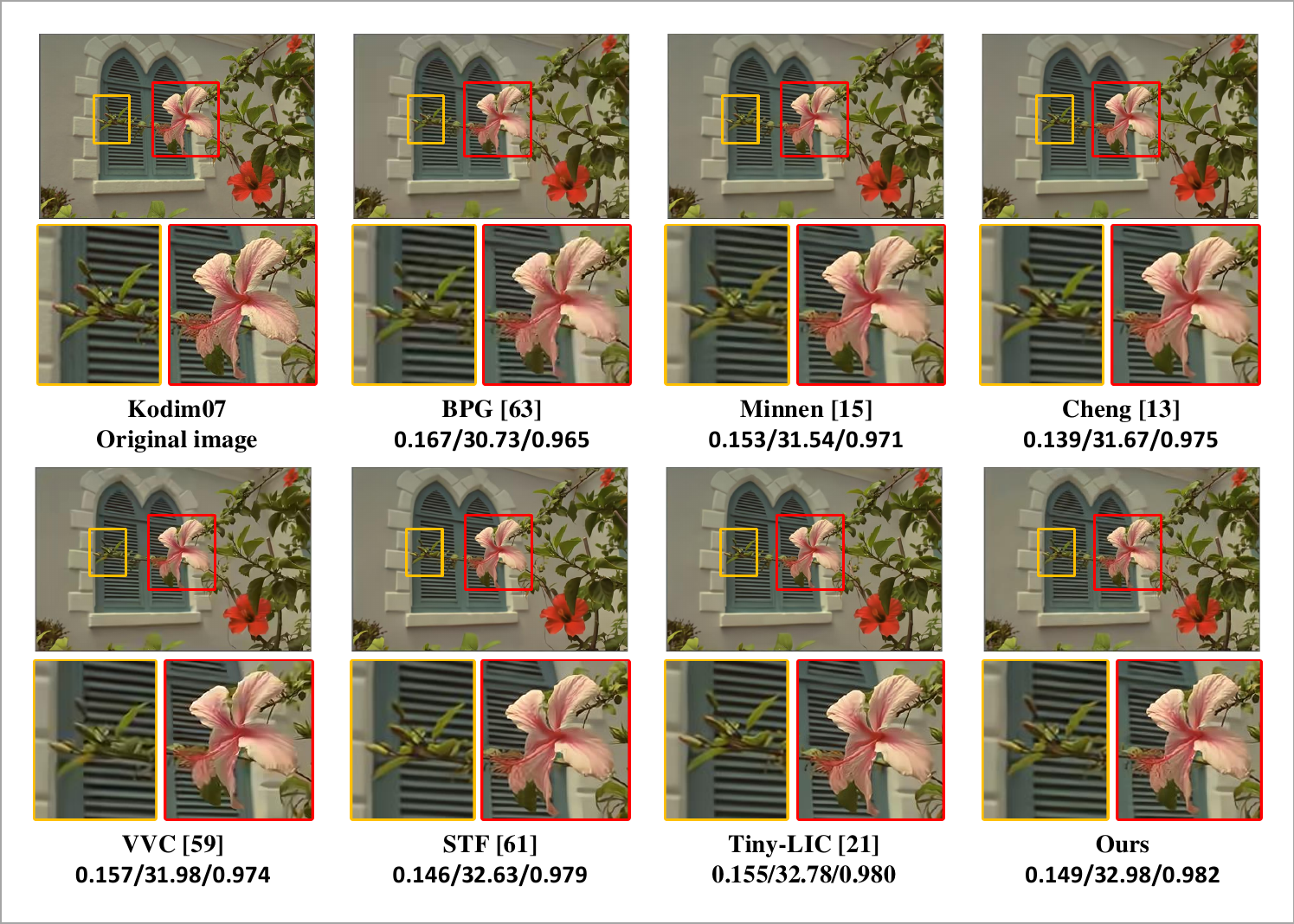}
	\caption{Visualization of “kodim07.png”for different methods. For each codec, we give the bitrate/PSNR/MS-SSIM.}
	\label{fig:subject-kodim07}
	\end{minipage}
\end{figure*}

\begin{figure}[htbp]
\begin{minipage}[b]{0.495\linewidth}
  \centering
  \includegraphics[width=1\linewidth]{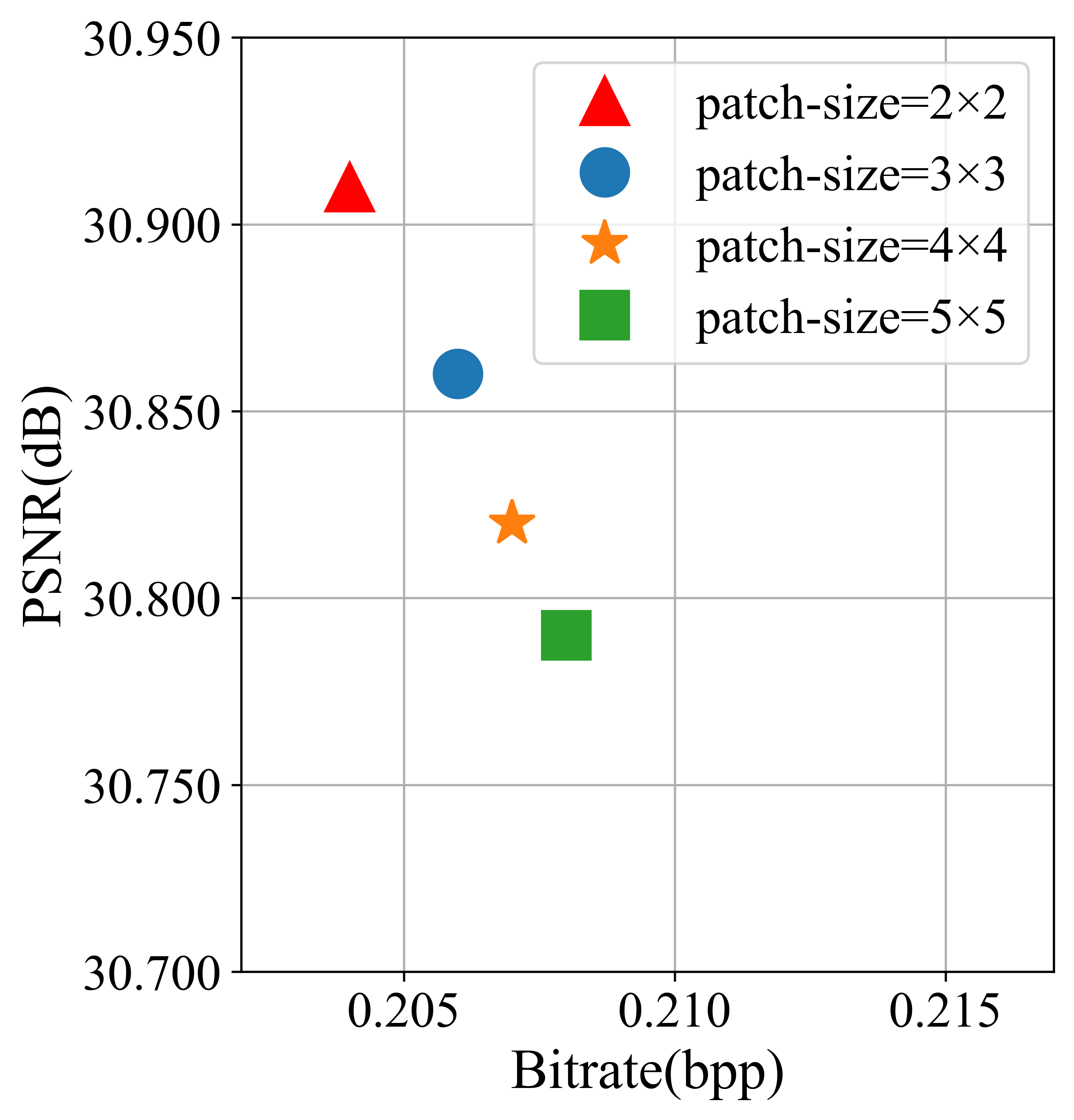}
  \centerline{(a) Kodak}\medskip
\end{minipage}
\begin{minipage}[b]{0.495\linewidth}
  \centering
  \includegraphics[width=1\linewidth]{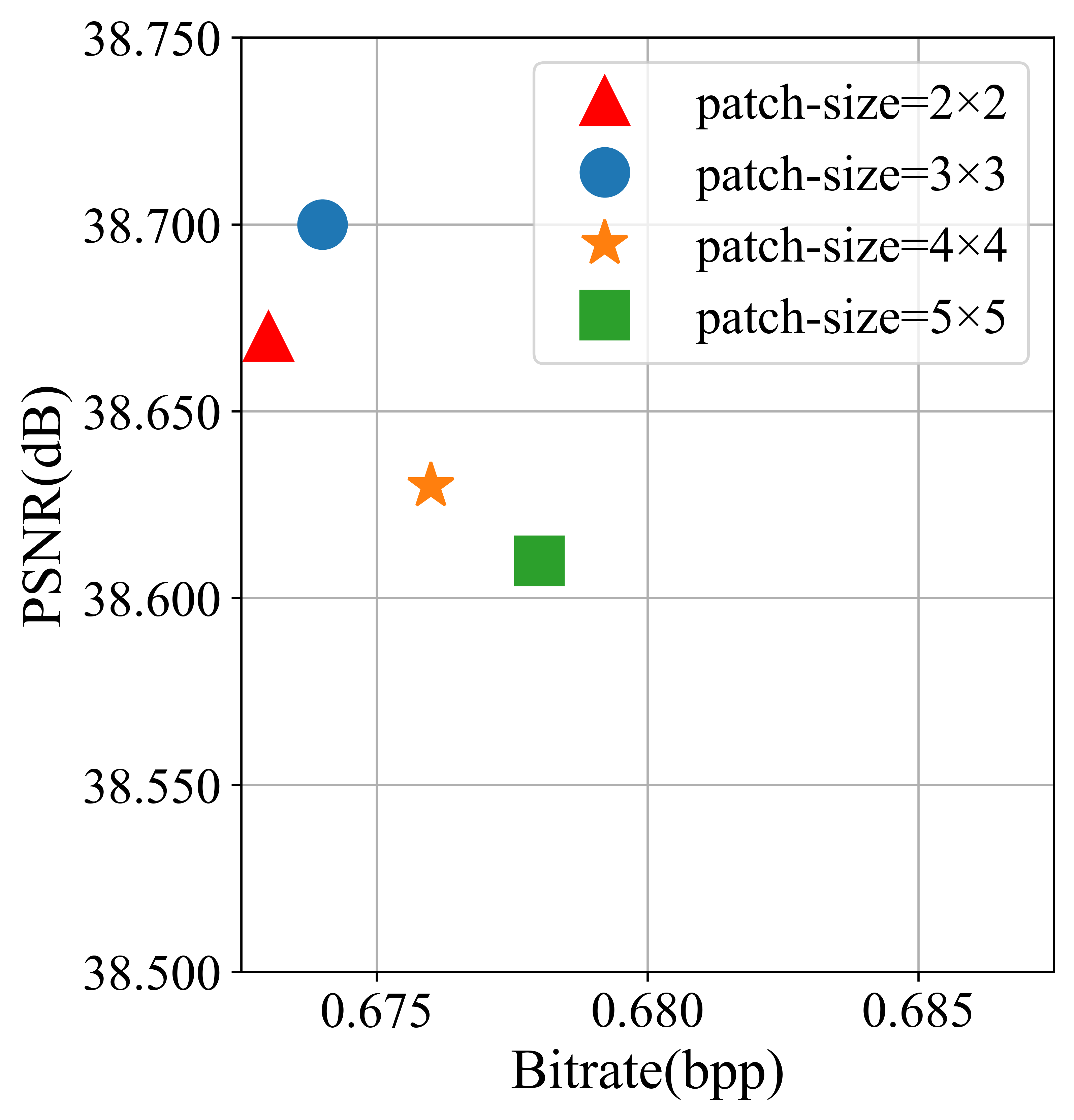}
  \centerline{(b) CLIC}\medskip
\end{minipage}
\caption{RD performance for different patch sizes $b \times b$.}
\label{fig:patch_size}
\end{figure}

\subsection{Complexity}
\textcolor{black}{We evaluated the number of network parameters and time complexity of various LIC methods on the Kodak dataset, as shown in Table \ref{tab:complexity}. Note that the encoding and decoding times of the parallel context models (LIC-TCM \cite{liu2023learned}, STF \cite{Zou_2022_CVPR}, Tiny-LIC \cite{lu2022high}) were lower than those of the autoregressive models (Minnen \cite{minnen2018joint}, Cheng \cite{cheng2020learned} and Xie \cite{xie2021enhanced}). Compared to the backbone \cite{lu2022high}, our method required more model parameters and longer encoding/decoding times. However, the incurred complexity was acceptable. Nevertheless, the computational complexity and encoding/decoding time of ``Proposed-Tiny" were similar to Tiny-LIC. At the same time, as shown in Table \ref{tab:bd-rate}, ``Proposed-Tiny" outperformed Tiny-LIC, indicating the efficiency of the proposed method. Table \ref{tab:complexity_module} gives the computational complexity of the proposed modules in terms of giga multiply-add operations per second (GMACs). It shows that the high complexity of the proposed method is mainly due to the FExM and FRM modules.}  

\subsection{Qualitative results}
\begin{figure*}[ht]
	\centering
	\begin{minipage}{\textwidth}
	\centering
	\includegraphics[width=0.99\textwidth]{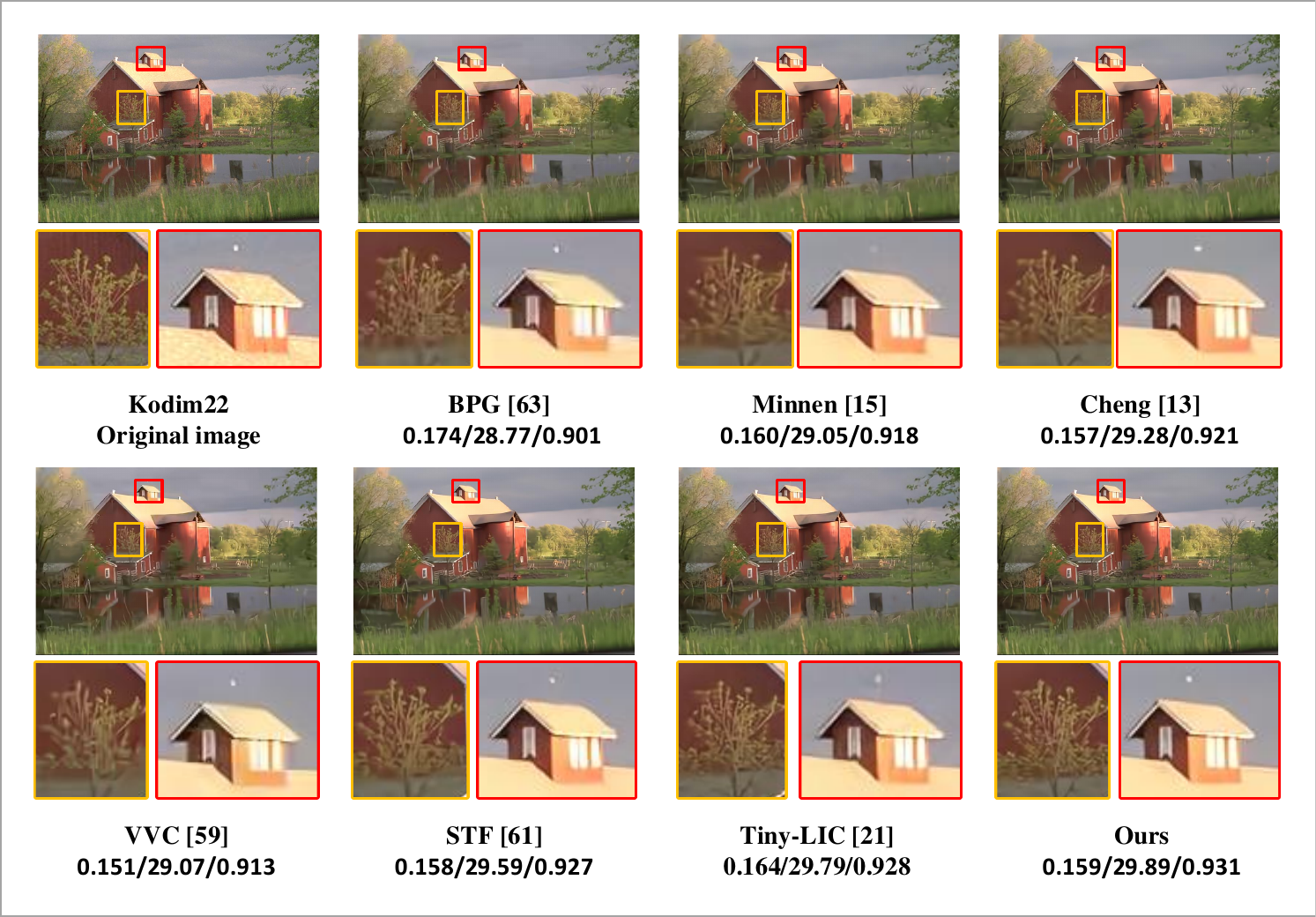}
	\caption{Visualization of “kodim22.png” for different methods. For each codec, we give the bitrate/PSNR/MS-SSIM.}
	\label{fig:subject-kodim22}
	\end{minipage}
\end{figure*}

In Figure \ref{fig:subject-kodim07} and Figure \ref{fig:subject-kodim22}, we illustrate the reconstruction quality of the codecs using Kodim07 and Kodim22 from the Kodak dataset as an example. The learning-based approaches outperformed H.266/VVC \cite{VVC}. Compared to \cite{lu2022high}, our method achieved better reconstruction quality at an even lower bitrate. In Figure \ref{fig:subject-kodim07}, it better preserved the flower details while in Figure \ref{fig:subject-kodim22} it better reconstructed the texture edges of the upper and lower right corner of the attic.

\subsection{Ablation study}
To demonstrate the effectiveness of components in the proposed method, we conducted ablation experiments.

\textbf{Performance of varying patch sizes.} Figure \ref{fig:patch_size} illustrates the impact of varying patch sizes on RD performance. Different patch sizes generate a variable number of sub-images, altering spatial intra-frame and temporal inter-frame correlations. Patch sizes were set to $2 \times 2$, $3 \times 3$, $4 \times 4$, and $5 \times 5$, and evaluations were performed on the Kodak and CLIC datasets. The results indicate that a patch size of $2 \times 2$ is optimal for the Kodak dataset, whereas a patch size of $3 \times 3$ is most effective for the CLIC dataset. This is attributable to the higher resolution of the CLIC dataset, where a larger patch size maintains spatial correlation without significant alteration and exploits temporal correlation to enhance RD performance effectively.

\textbf{Performance of various components.} \textcolor{black}{To validate the efficiency of various components in the proposed method, we conducted ablation experiments on the Kodak dataset, as shown in Table \ref{tab:ablation}. To calculate the BD-rate, we used four $\lambda$s (0.0035, 0.0067, 0.013, and 0.0483) and reported the bitrate and PSNR for different modules at $\lambda$ values of 0.0035 and 0.0483. Adding FExM and FRM to the baseline method, which uses the method in \cite{lu2022high} as a backbone, increased the PSNR by 0.1 dB at a similar bitrate. Further adding FEnM and QECM yielded improvements of about 0.1 dB each. It can be observed that the improvement contributed by the FEnM network with the loss function $\mathcal{L}_{FE}$ was more significant than that contributed by the FEnM network without $\mathcal{L}_{FE}$. Our approach achieved a BD-rate reduction of 3.6\% over the baseline model. The results demonstrate that the proposed modules can significantly enhance the RD performance, especially at higher bitrates.}

\begin{figure}[ht]
	\centering
	\includegraphics[width=0.9\linewidth]{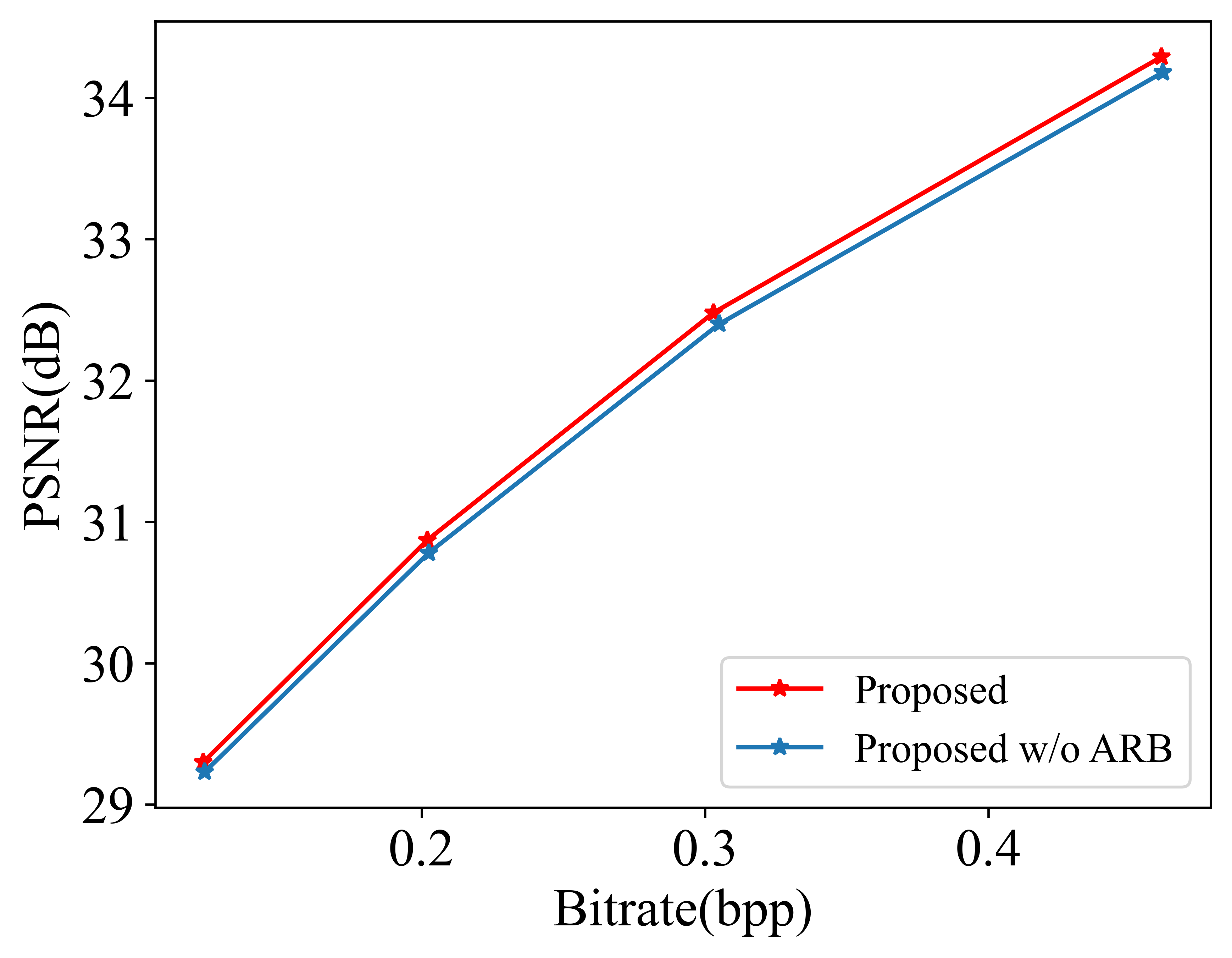}
	\caption{Effect of the feature attention refinement block.}
	\label{fig:no_ARB}
\end{figure}

\textbf{Performance of attention refinement block.} Figure \ref{fig:no_ARB} displays the RD curves of the proposed method and the method without ARB on the Kodak dataset. ARB extracted features more effectively, improving the RD performance.

\begin{figure}[ht]
	\centering
	\includegraphics[width=0.9\linewidth]{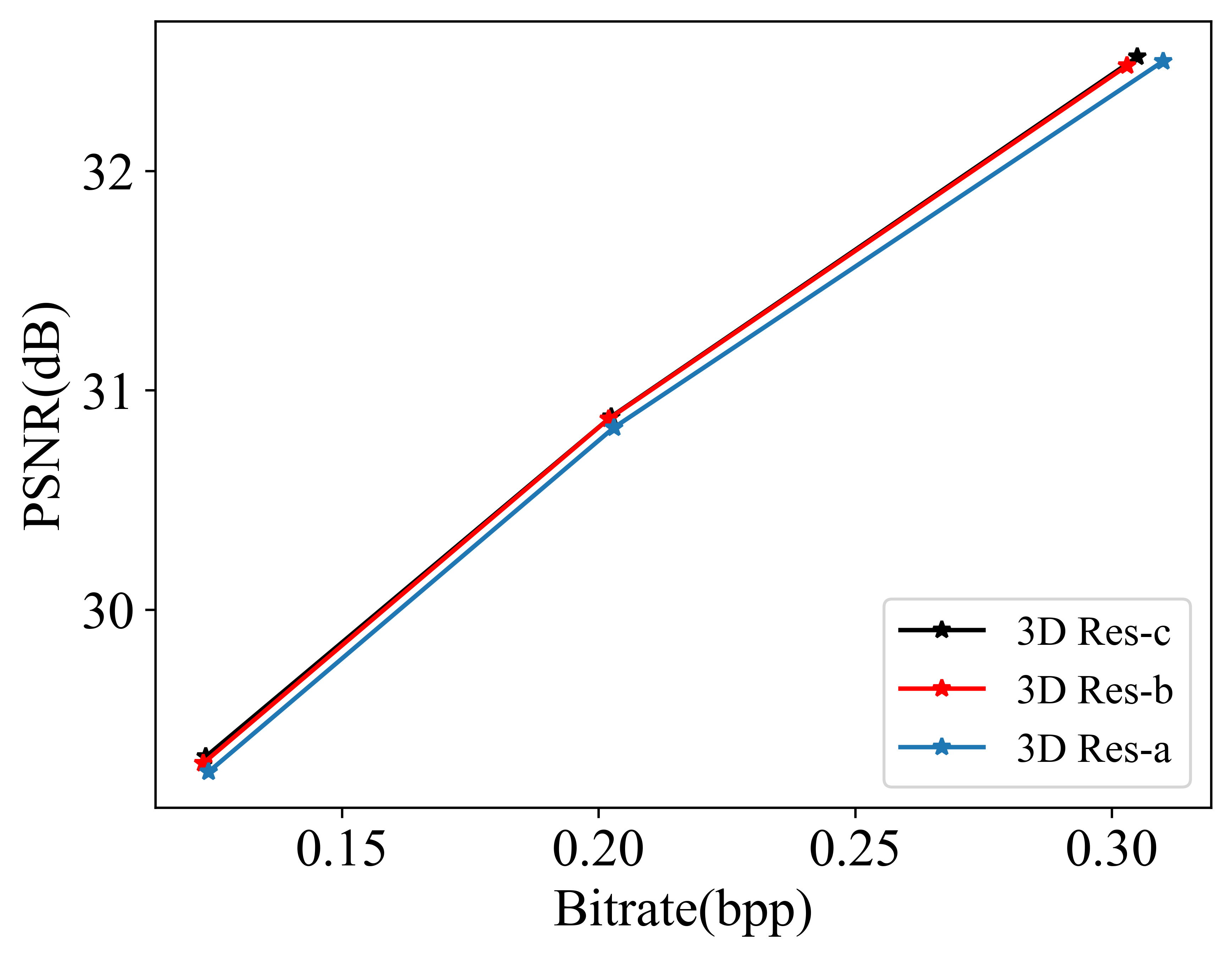}
	\caption{Effect of the number of levels in 3D Res.}
	\label{fig:3D_Res_rd}
\end{figure}

\textbf{Performance of different concatenated residual modules.} Figure \ref{fig:3D_Res_rd} illustrates the impact of the number of levels in 3D Res. Increasing the number of levels in 3D Res improved the RD performance to a certain extent, as a larger receptive field can extract more effective features. However, the improvement was limited. The performance of the two-level 3D Res was similar to that of the three-level 3D Res. Therefore, considering the trade-off between performance and complexity, we chose the two-level 3D Res to construct our model.

\subsection{Flexibility analysis}
\begin{figure}[ht]
\begin{minipage}[b]{0.495\linewidth}
  \centering
  \includegraphics[width=1\linewidth]{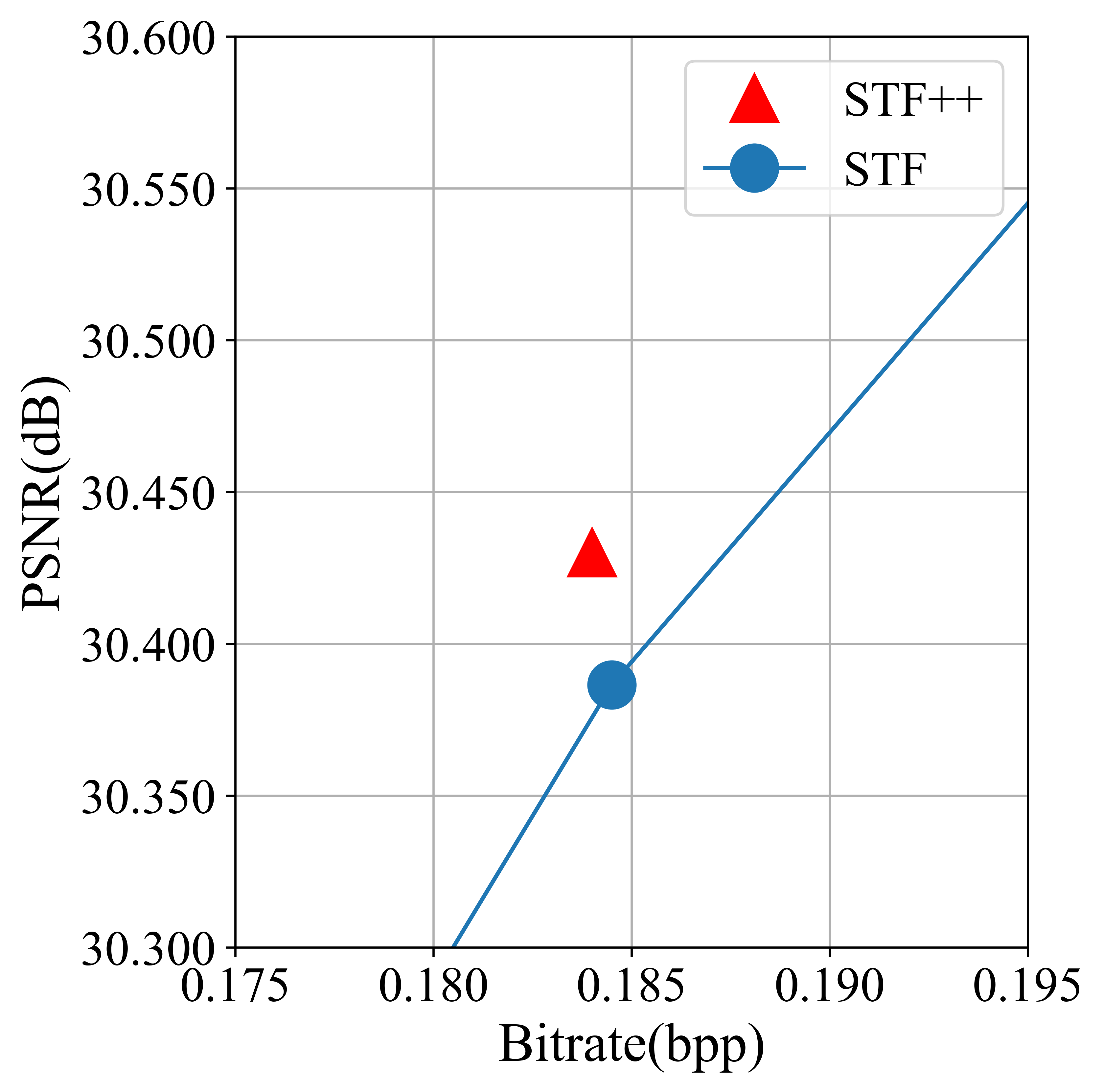}
  \centerline{(a) $\lambda$ =0.0035}\medskip
\end{minipage}
\begin{minipage}[b]{0.495\linewidth}
  \centering
  \includegraphics[width=1\linewidth]{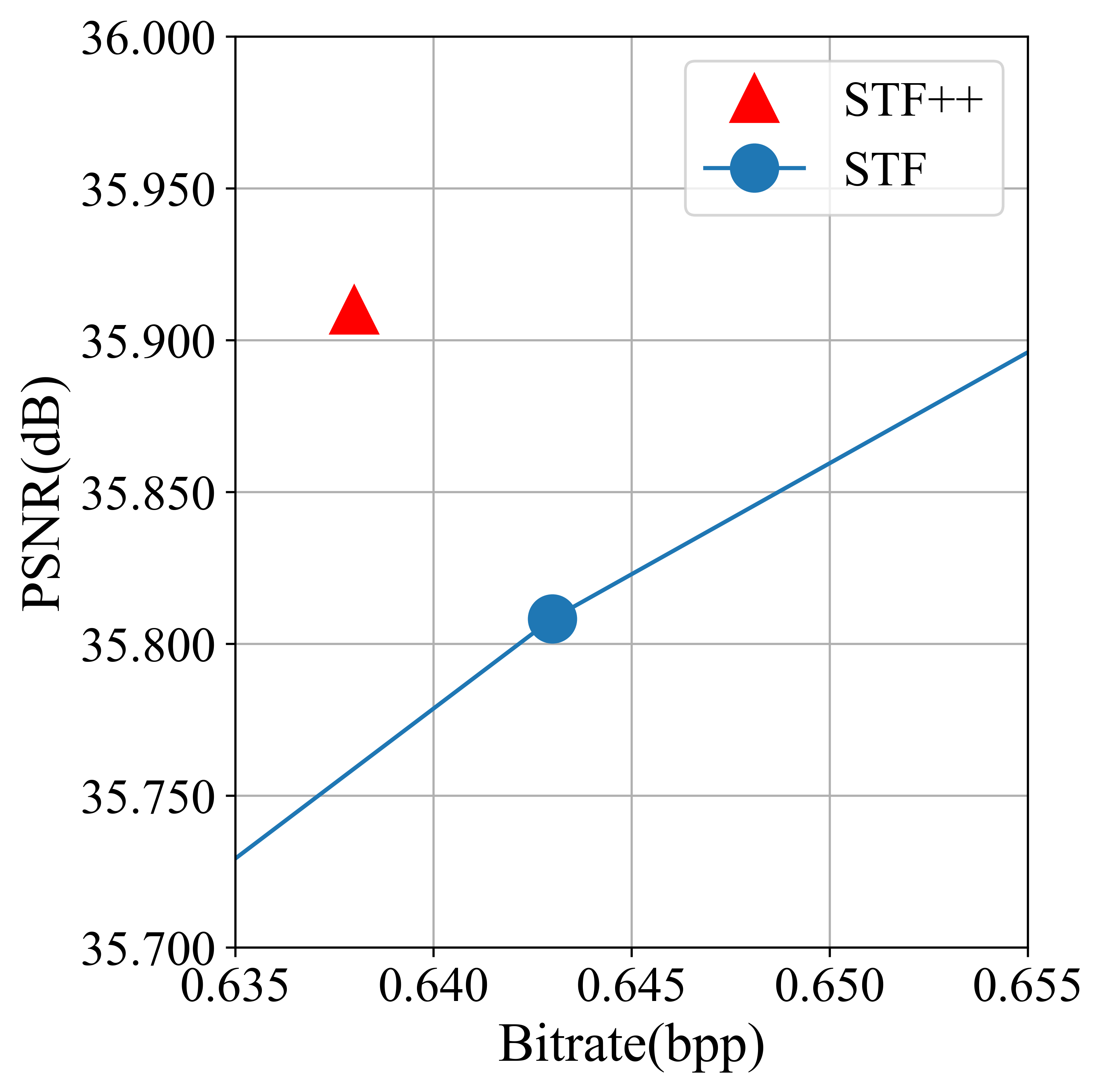}
  \centerline{(b) $\lambda$ =0.025}\medskip
\end{minipage}
\caption{Performance of the proposed method when applied to STF \cite{Zou_2022_CVPR}. STF++ is STF enhanced with the proposed four modules. }
\label{fig:stf++}
\end{figure}

In principle, the proposed method can be easily combined with any state-of-the-art LIC scheme. In the previous sections, we used the method in \cite{lu2022high} as the backbone network. To assess the flexibility of our method, we replace in this section the method in \cite{lu2022high} with STF \cite{Zou_2022_CVPR} in the testing phase. Figure \ref{fig:stf++} shows the results. ``STF++" represents the proposed method when combined with STF. The proposed method significantly enhanced the performance of the backbone model, particularly exhibiting substantial improvements at high bitrates. This observation underscores the high flexibility of the proposed method.

\section{Conclusion}
\textcolor{black}{We proposed a learned image compression method that leverages feature extraction, feature refinement, feature enhancement, and quantization error compensation. The feature extraction module shuffles the pixels in the input image, splits the resulting image into sub-images, and uses cascaded convolutional layers to extract features from each sub-image. The feature refinement module stacks the extracted features and uses concatenated 3D Res blocks to capture correlations across channels, within sub-images (intra-sub-image correlation), and across sub-images (inter-sub-image correlation). The feature enhancement module mitigates information loss in the decoded features. The quantization error compensation module alleviates the mismatch between training and testing. The proposed modules can be flexibly integrated into existing LIC methods as plug-and-play components. Experimental results on the Kodak and CLIC datasets showed that our modules improved the RD performance of state-of-the-art LIC methods. In the future, we will explore smaller lightweight networks to further reduce the computational complexity of our approach. In addition, we will extend our ideas to other tasks for 3D engineering applications \cite{li20233d}.}

\bibliographystyle{IEEEtran}
\bibliography{ski}


\end{document}